\documentclass[preprint]{elsarticle}

\usepackage{lineno,hyperref}
\modulolinenumbers[5]

\usepackage{mathtools}
\usepackage{amsmath}
\usepackage{float}
\graphicspath{{figures/}}
\usepackage{subcaption} 
\usepackage{amssymb}
\usepackage{todonotes}
\usepackage{sidecap}

\usepackage[toc,page]{appendix} 

\usepackage{tabularx}

\usepackage[margin=3cm]{geometry}

\makeatletter
\def\ps@pprintTitle{%
   \let\@oddhead\@empty
   \let\@evenhead\@empty
   \let\@oddfoot\@empty
   \let\@evenfoot\@oddfoot
}
\makeatother

\begin{document}

\begin{frontmatter}

\title{Modelling realistic 3D deformations of simple epithelia in dynamic homeostasis}

\author[unimelb,sbl]{Domenic P.J. Germano}

\author[unimelb,sbl]{Stuart T. Johnston}
\author[sbl,med]{Edmund J. Crampin}
\author[unimelb]{James M. Osborne\corref{mycorrespondingauthor}}
\cortext[mycorrespondingauthor]{Corresponding author}
\ead{jmosborne@unimelb.edu.au}

\address[unimelb]{School of Mathematics and Statistics, University of Melbourne, Parkville, Victoria 3010, Australia }
\address[sbl]{Systems Biology Laboratory, School of Mathematics and Statistics, and Department of Biomedical Engineering, The University of Melbourne, Parkville, Victoria 3010, Australia
}
\address[med]{School of Medicine, Faculty of Medicine, Dentistry and Health Sciences, The University of Melbourne, Parkville, Victoria 3010, Australia}


\begin{abstract}
The maintenance of tissue and organ structures during dynamic homeostasis is often not well understood. In order for a system to be stable, cell renewal, cell migration and cell death must be finely balanced. Moreover, a tissue’s shape must remain relatively unchanged. Simple epithelial tissues occur in various structures throughout the body, such as the endothelium, mesothelium, linings of the lungs, saliva and thyroid glands, and gastrointestinal tract. Despite the prevalence of simple epithelial tissues, there are few models which accurately describe how these tissues maintain a stable structure.

Here, we present a novel, 3D, deformable, multilayer, cell-centre model of a simple epithelium. Cell movement is governed by the minimisation of a bending potential across the epithelium, cell-cell adhesion, and viscous effects. We show that the model is capable of maintaining a consistent tissue structure while undergoing self renewal. We also demonstrate the model's robustness under tissue renewal, cell migration and cell removal. The model presented here is a valuable advancement towards the  modelling of tissues and organs with complex and generalised structures.
\end{abstract}

\end{frontmatter}

\newpage

\section{Introduction}
The body consists of four basic tissue types: epithelial, connective, muscular and nervous \cite{kahle1976color}. The epithelium line the internal and external surfaces of the body, as well as cavities, and some organs and glands. These epithelia are the functioning  components of the tissues, capable of performing various tasks, including protection, secretion, absorption, excretion, filtration, diffusion and sensation, depending on tissue location.
Epithelial tissues have a high potential for  malignancies, in the form of cancers, as well as many other diseases, such as asthma, cardiac disease, and many viral induced diseases. Therefore, to better understand how disease develops, a clear understanding towards how healthy epithelia are maintained in homeostasis is first required.

Epithelia can be categorised in a number of ways, with one possible way being the organisation, shape and function of the constituent cells. In terms of cell shape, epithelial cells can be: \textit{squamous}, cells with a width-to-height ratio greater than 1; \textit{cuboidal}, cells with a width-to-height ratio approximately equal to 1; or \textit{columnar}, cells with a width-to-height ratio less than 1.
Tissue organisation is categorised as either \textit{simple}, if the epithelium contains only a single layer of epithelial cells; \textit{stratified}, if the epithelium contains two or more layers of epithelial cells; or \textit{pseudostratified}, when the epithelium is made up of a single layer of columnar cells of non uniform width along the long axis \cite{kahle1976color}.
Simple epithelial tissues are common throughout the human body, and occur in various geometric structures, depending on the constituent cells' shape. See Table \ref{Tab:Epithelium_Types} for a schematic of simple epithelial tissue types, and some examples of their locations within the body. 

\begin{table}[h!]
\centering
\begin{tabular}{ c | c | c} 
 Epithelium type  & Location & Reference \\
 \hline
 \begin{minipage}{0.4\linewidth}
  \vspace{0.2cm}
\centering Simple squamous \newline
\includegraphics[width = 0.6\textwidth]{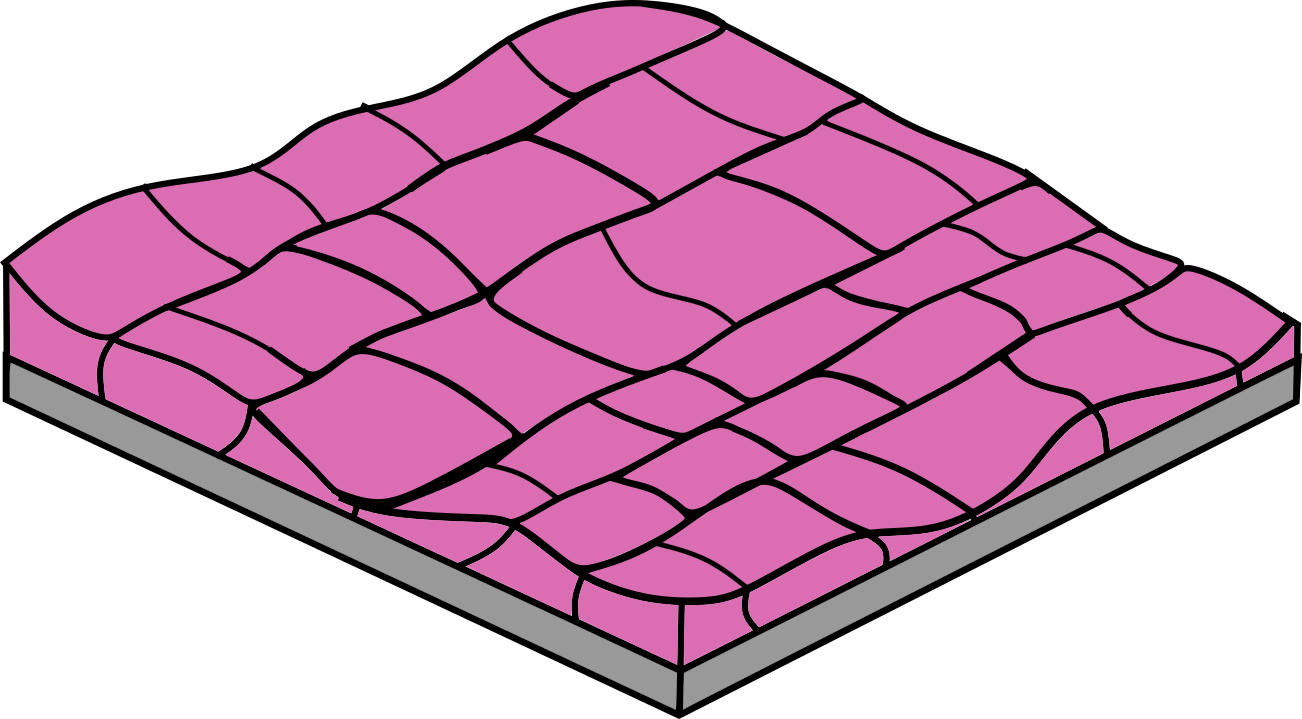}
  \vspace{0.2cm}
\end{minipage}
 & \begin{minipage}{0.4\linewidth}
\begin{description}
\setlength\itemsep{0.0em}
    \item Endothelium (capillary walls)
    \item Lung alveoli
    \item Mesothelium (peritoneum)
\end{description}
\end{minipage}
& \begin{minipage}{0.15\linewidth}
\begin{description}
\setlength\itemsep{0.0em}
    \item \cite{stolz2015unwrapping}
    \item \cite{bonastre2016cell}
    \item \cite{hiriart2019mesothelium}
\end{description}
\end{minipage}\\
\hline
 \begin{minipage}{0.4\linewidth}
\centering Simple cuboidal \newline
\includegraphics[width = 0.6\textwidth]{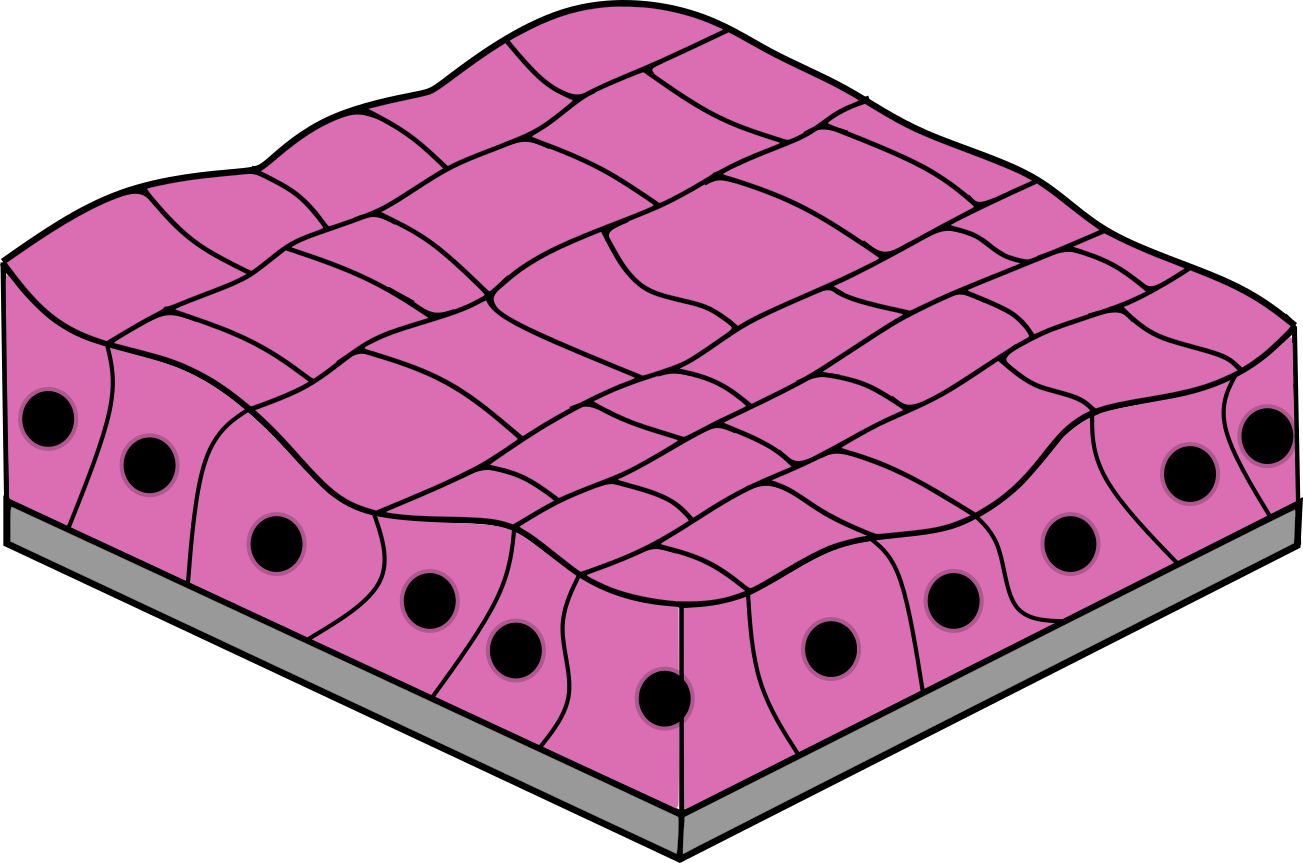}
\end{minipage}
 & \begin{minipage}{0.4\linewidth}
  \vspace{0.2cm}
\begin{description}
\setlength\itemsep{0.0em}
    \item Ovary surface
    \item Renal tube lining
    \item Linings of the lungs
    \item Saliva glands
    \item Eyes
    \item Thyroid glands
\end{description}
  \vspace{0.2cm}
\end{minipage}
& \begin{minipage}{0.15\linewidth}
\begin{description}
\setlength\itemsep{0.0em}
    \item \cite{katabuchi2003cell}
    \item \cite{thakur2019determination}
    \item \cite{hermans1999lung}
    \item \cite{de2017overview}
    \item \cite{frost2014autophagy}
    \item \cite{balasubramanian2020anatomy}
\end{description}
\end{minipage}\\
\hline
 \begin{minipage}{0.4\linewidth}
 \vspace{0.2cm}
\centering Simple Columnar \newline
\includegraphics[width = 0.6\textwidth]{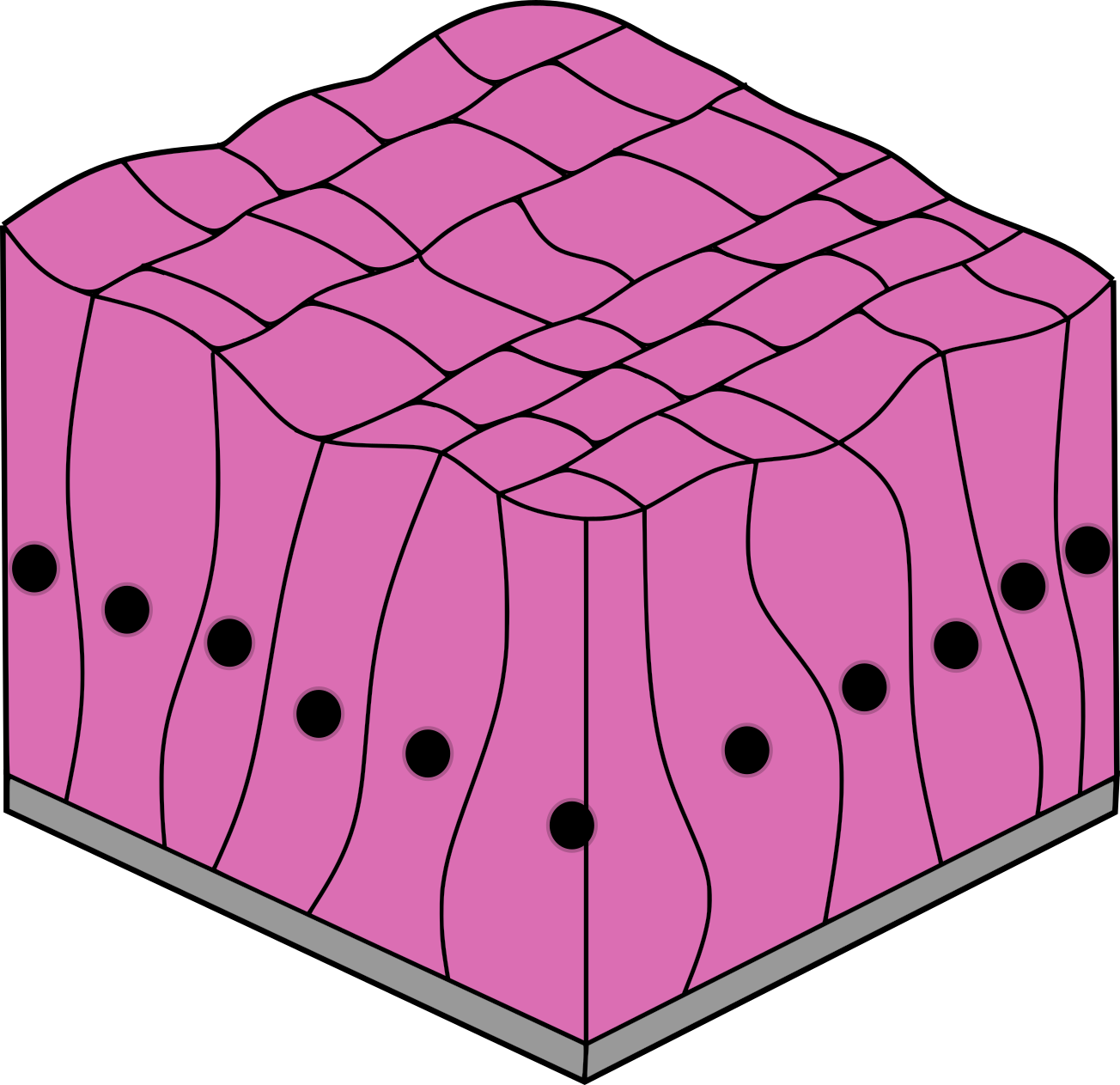}
\end{minipage}
 & \begin{minipage}{0.4\linewidth}
\begin{description}
\setlength\itemsep{0.0em}
    \item Gastrointestinal tract
    \item Endocervix
    \item Fallopian tubes
\end{description}
\end{minipage}
& \begin{minipage}{0.15\linewidth}
\begin{description}
\setlength\itemsep{0.0em}
    \item \cite{reed2009review}
    \item \cite{yi2013genital}
    \item \cite{wira2005epithelial}
\end{description}
\end{minipage}\\
\end{tabular}
\caption{\label{Tab:Epithelium_Types} Schematic of simple epithelial tissues, and their locations within the body.}
\end{table}

 Simple epithelia reside on a basement membrane, supported by the connected stroma tissue \cite{kahle1976color}. They consist of self-renewing cells, which vary in cell cycle duration depending on the tissue's location within the body. Despite the frequency of simple epithelial tissues throughout the body, the properties leading to their particular geometric structures are not well understood.
 To better understand their function, many different mathematical models of simple epithelia exist. For example, models can incorporate cells as discrete interacting agents. This is typically done on a 2D planar geometry \cite{an2008introduction}, or a fixed 3D geometry \cite{dunn2013computational}. While valuable and insightful, these approaches provide limited information toward the geometric structure of the epithelium. Another approach is to model the epithelium as a continuum string (2D) \cite{almet2021role} or sheet (3D) \cite{hannezo2011instabilities} of cells. However, the insights gained into the geometric structure from deformable continuum models comes with a loss of information into cell topology.
 
 In this work, we present a novel model of realistic 3D deformations of simple epithelium. The aim of this model is to be able to accurately incorporate both realistic tissue geometries and cell topology. The model is generalised, with the goal of being customisable, based on the modelling applications.



\section{Model of simple epithelium}
\label{sec:model}
Here, we present a mathematical model of a simple epithelium. We model the tissue using a multicellular model, where cells are represented by their centres, which are free to move in space \cite{meineke2001cell}. We use multiple layers of cells: an epithelial layer, in which cells are able to proliferate, and a stromal layer, which supports the epithelium, in which cells are differentiated and do not proliferate.

\subsection{Biomechanical model}
We use a lattice-free, cell-centred model, to describe the biomechanical forces experienced by cells, with $\mathbf{r}_i$ denoting the position of cell centre $i$ \cite{osborne2017comparing}. We denote the net force on a given cell $i$, $\mathbf{F}_{i}^{\text{Net}}$. Here, we include forces due to neighbouring interactions, $\mathbf{F}_{i}^{\text{Interaction}}$, viscous forces, $\mathbf{F}_{i}^{\text{Viscous}}$ and lastly the bending force, $\mathbf{F}_{i}^{\text{Bending}}$. The net force is therefore
\begin{align}
    \mathbf{F}_{i}^{\text{Net}} = \mathbf{F}_{i}^{\text{Interaction}} + \mathbf{F}_{i}^{\text{Viscous}} + \mathbf{F}_{i}^{\text{Bending}}, \quad \forall i.
\end{align}
We assume that cell motion is over-damped, due to the highly viscous cellular environment \cite{dallon2004cellular}. Ignoring inertial terms, the net force on any given cell-centre is zero, i.e. $\mathbf{F}_{i}^{\text{Net}} = \mathbf{0}$. We model elastic neighbour interactions via a spring potential energy, $ {U}^{\text{Interaction}} $. We first write the interaction potential for a single cell $j$ as $P_j$:
\begin{align}
P_j = \sum_{n \in N_j} \frac{1}{2} p_{jn} \left(  s_{jn} - \vert \mathbf{r}_{jn} \vert \right)^2,
\end{align}
where $N_j$ is the set of the first-order neighbours, given by the Delaunay triangulation, $0 \leq p_{jn} \leq 1$ describes how readily cells $j$ and $n$ interact, $\mathbf{r}_{jn}$ is the displacement between cell-centres $j$ to $n$, and $s_{jn}$ is the rest separation between cells $j$ and $n$ in the absence of other external forces. The interaction potential of the tissue is
\begin{align}
    {U}^{\text{Interaction}} =  \sum_{\forall j} k_{j} P_j,
\end{align}
where $k_{j} \geq 0$ is the spring constant and denotes how readily cell $j$ adheres to its neighbours. The interaction force experienced by cell $i$ is then defined as the minimisation of $ {U}^{\text{Interaction}} $:
\begin{align}
    \mathbf{F}_{i}^{\text{Interaction}} = - \nabla_i  {U}^{\text{Interaction}}, \quad \forall i.
\end{align}
We note that this is equivalent to  a conventional spring force in the form $\mathbf{F}_i = \sum_{n \in N
_i} \kappa_{in} \left(  \vert \mathbf{r}_{in} \vert - s_{in} \right)  \hat{\mathbf{r}}_{in} $, where $\kappa_{in}$ is the spring constant between cells $i$ and $n$.
The viscous force simply opposes the direction of motion
\begin{align}
     \mathbf{F}_{i}^{\text{Viscous}} = - \nu_i \frac{d \mathbf{r}_i}{dt}, \quad \forall i,
\end{align}
where $\nu_i > 0$ is the drag coefficient of cell $i$.
\subsection{Bending force}
\begin{figure}[H]
\centering
\begin{subfigure}[b]{0.99\textwidth}
\centering
\includegraphics[width = 0.475\textwidth]{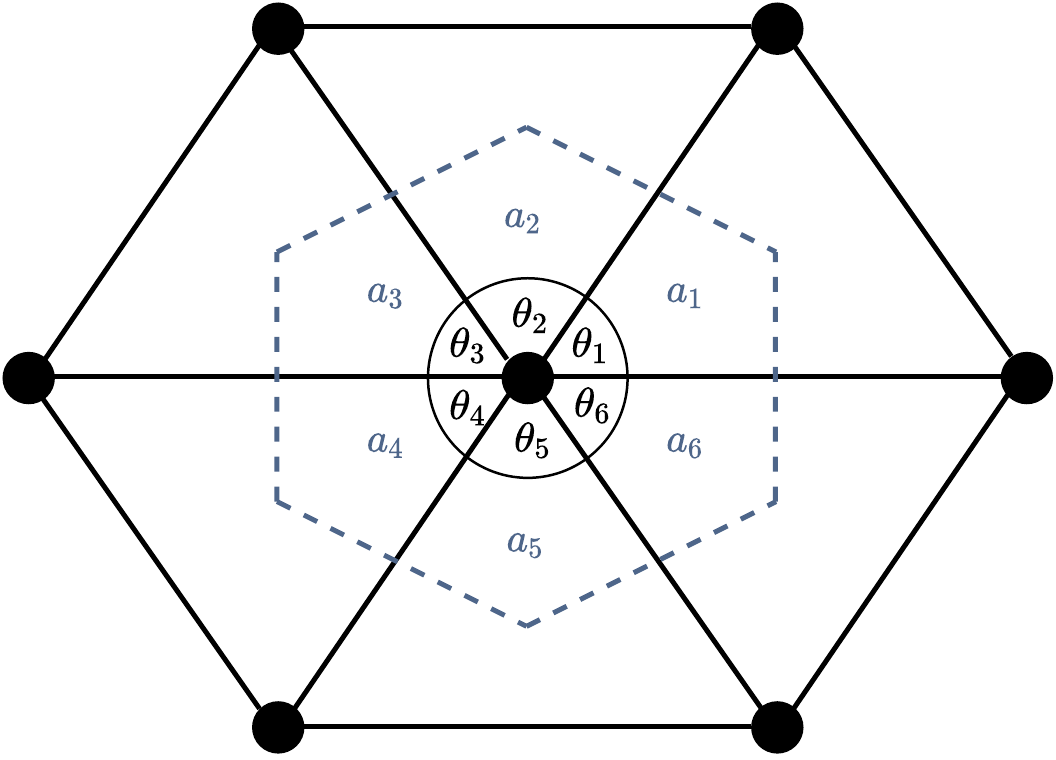}
\caption{\label{fig:dGc_element} }
\end{subfigure}

\begin{subfigure}[b]{0.32\textwidth}
\centering\includegraphics[width=0.7\textwidth]{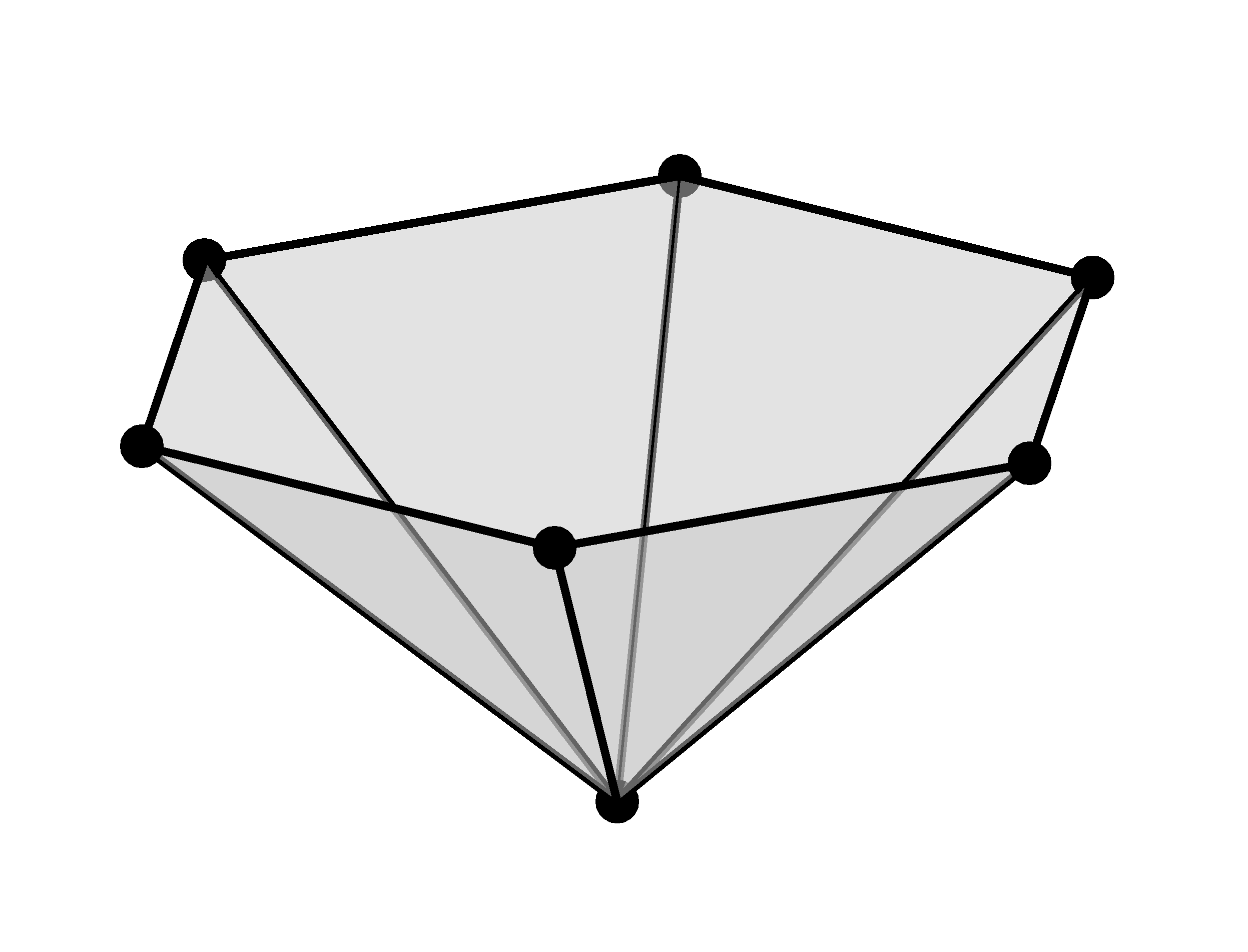}
\caption{$G_j>0$ \label{fig:parabaloid}}
\end{subfigure}
\begin{subfigure}[b]{0.32\textwidth}
\centering\includegraphics[width=0.7\textwidth]{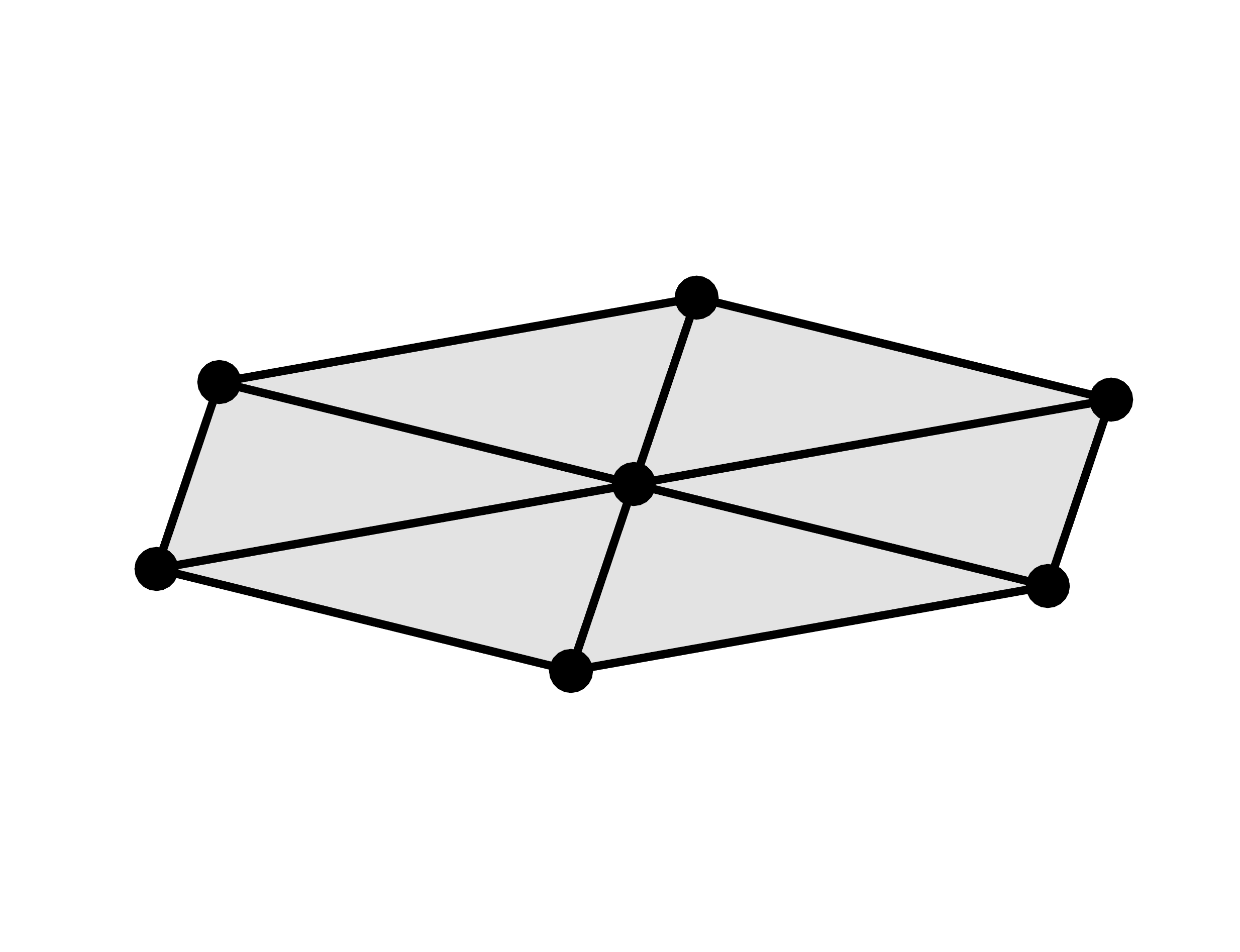}
\caption{$G_j=0$\label{fig:flat}}
\end{subfigure}
\begin{subfigure}[b]{0.32\textwidth}
\centering\includegraphics[width=0.7\textwidth]{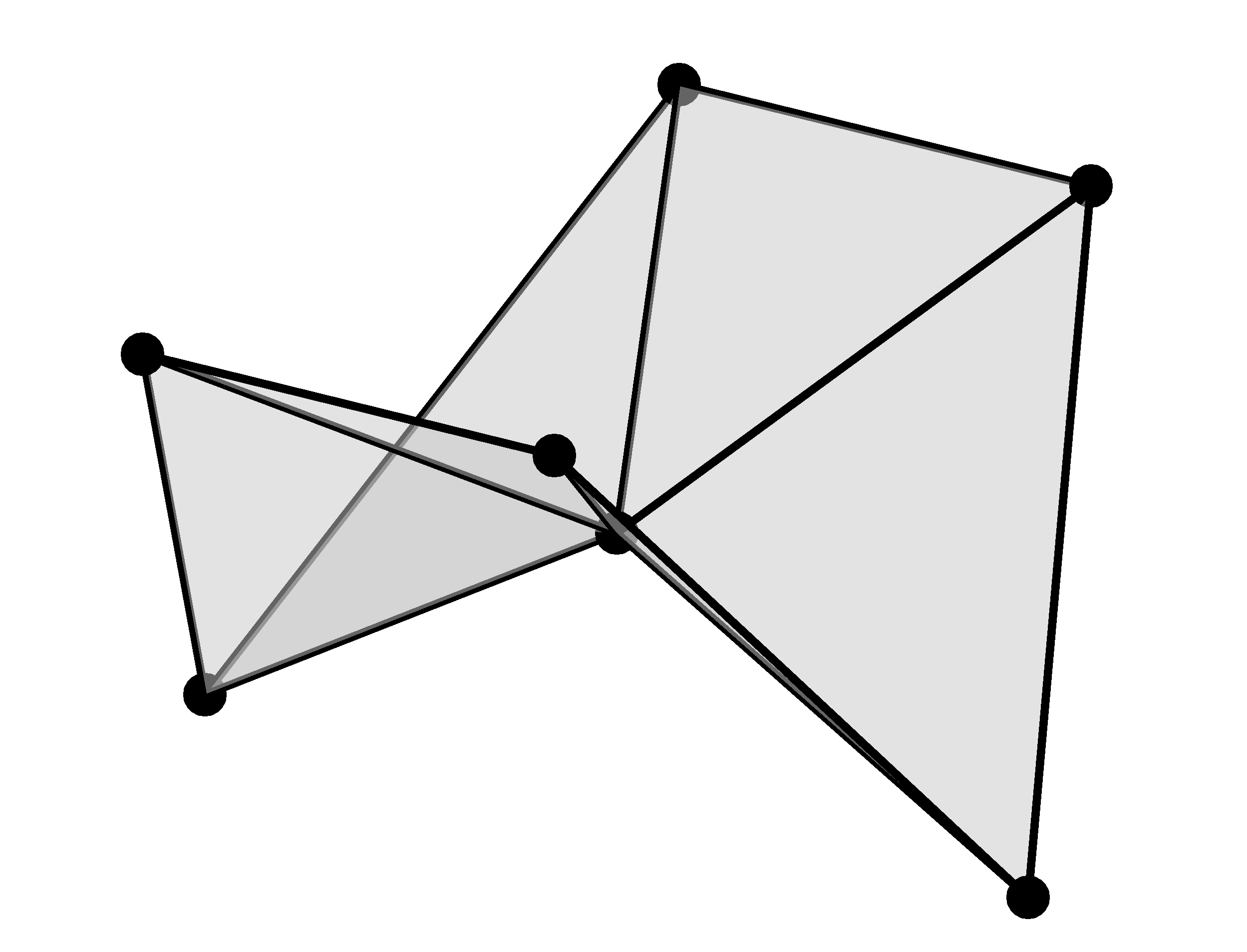}
\caption{$G_j<0$\label{fig:hyperbolic}}
\end{subfigure}

\captionsetup{subrefformat=parens} 
\caption{ 
\subref{fig:dGc_element} Local element used to calculate the discrete Gaussian curvature.  Possible element configurations, for different possible values of $G_j$: an elliptical element shown in \subref{fig:parabaloid},  a flat element shown in \subref{fig:flat} and  hyperbolic element shown in \subref{fig:hyperbolic}.}
\label{fig:element_configurations}
\end{figure}

If we want to examine the shape of a given tissue, we first require a measure of  ``shape''. A common technique to quantify the shape of a triangulated surface is the discrete Gaussian curvature \cite{xu2009discrete}, defined locally at each vertex $j$ in the surface $S$ as:
\begin{align}
    G_j  &= \frac{ 2\pi - \phi_j }{A_j}, \quad \forall j,
\end{align}
where $ \phi_j = \sum_{n \in T_j} \theta_n $ is the angle sum at $j$, $T_j$ are the set of vertices that share an edge with vertex $j$, $\theta_n$ is the angle of the $n^{\text{th}}$ triangular-element at vertex $j$, and $A_j=\sum_{n \in T_j} a_n$ is the area contribution of the element that contains vertex $j$ on the surface $S$, as shown by Figure \ref{fig:dGc_element} where $a_n$ is calculated as one third of the area of triangular-element $n$.


For a hexagonally-packed set of points, there are three possible configurations for the element shown in Figure \ref{fig:dGc_element}. We depict these configurations in Figure \ref{fig:parabaloid}--\ref{fig:hyperbolic}. If we consider the element shown in Figure \ref{fig:dGc_element} as a single element, we can say the element is elliptic if the discrete Gaussian Curvature $G_j>0$, flat if $G_j=0$, and hyperbolic if $G_j < 0$.

Since our model defines cell connectivity within the tissue via a Delaunay triangulation, we utilise this triangulation and consider only the epithelial monolayer  to defined a surface $S^E$. Using this surface, we classify the global shape of the tissue through a bending potential, $U^{\text{Bending}}$, as:
\begin{align}
    U^{\text{Bending}} &=  \sum_{\forall j} \beta_j \vert G_j \vert ^{\alpha},\label{eq:bending_potential} 
\end{align}
where $\alpha >0$ is the bending exponent and $\beta_j$ is the magnitude of the bending potential at cell $j$ with:
\begin{align}
\beta_j = \begin{cases}
     \beta, &\text{if } j\in N^E,\\
     0, &\text{otherwise},
    \end{cases}
\end{align}
where $\beta \in \mathbb{R}$ and $ N^E$ is the set of epithelial cells within the tissue. We now minimise the bending potential, which will result in the epithelial monolayer surface deforming to minimise the global curvature. In this instance, minimising the bending potential results in a flat epithelial monolayer surface, $S^E$. We therefore write the bending force on epithelial cell centre $i$ as:
\begin{align}
    \mathbf{F}^{\text{Bending}}_{i} &= - \nabla_i U^{\text{Bending}},  \quad \forall i.
\end{align}
For a detailed explanation of the exact form of the gradient of the bending force, see  \ref{sec:eq_of_motion}. We can now write the equations of motion as:
\begin{align}
\nu_i \frac{d \mathbf{r}_i}{dt} = -\nabla_i \left( \sum_{\forall j} k_j P_j +  \sum_{\forall j} \beta_j \vert G_j \vert^\alpha \right), \quad \forall i. \label{eq:motion_eq_1}
\end{align}
\subsection{Tissue geometry}
\begin{figure}[H]
\centering
\includegraphics[width = 0.4\textwidth]{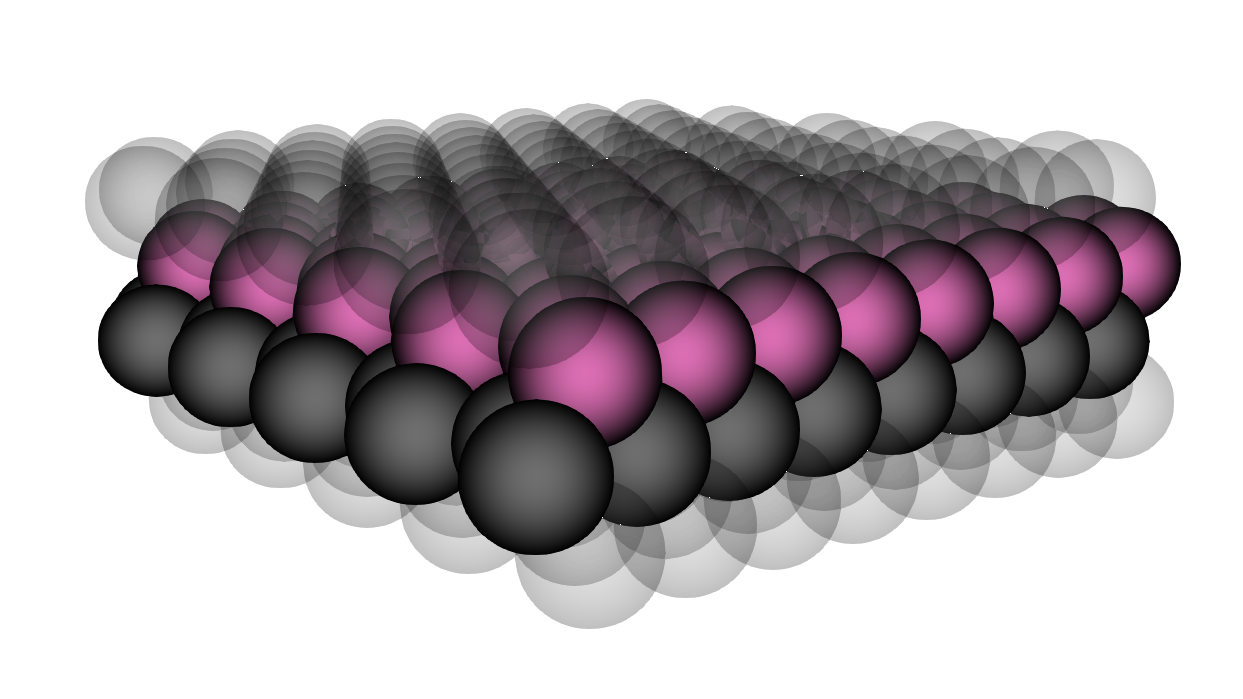}
\caption{\label{fig:initial_tissue} Initial tissue set up of size $\left[ N_x, N_y, N_z, N_g\right] = \left[8,10,2,1\right]$. Epithelial cells are shown in pink, stromal cells in dark grey and ghost nodes are light (transparent) grey.}
\end{figure}

To best replicate \textit{in vitro} and \textit{in vivo} conditions, we simulate the tissue on a periodic domain, to mimic the behaviour of a much larger tissue
\cite{germano2020mathematical, fletcher2013implementing, MILLER2021110807}. We impose periodic boundaries in the $x$ and $y$ axes. We allow the $z$ axis to be a free boundary at both the top and bottom. Below, the tissue is supported by the extracellular matrix, and above, the tissue is exposed to the open lumen. 
However, since our simulation method uses a Delaunay triangulation to define cell connectivity, we incorporate passive ghost nodes above and bellow the tissue to prevent long edges forming between cells which cannot possibly be connected to one another \cite{osborne2017comparing}. An example of the tissue configuration with ghost nodes is shown in Figure \ref{fig:initial_tissue}.

The cell packing we use to initialise the tissue is called hexagonal close packed. With a hexagonal close packed tissue, cells are initially separated by their rest length, and therefore artefacts due to the tissue relaxing are not present. We specify the tissue size as $\left[ N_x, N_y, N_z, N_g\right]$ where $N_x$ is the number of cells packed along the $x$ axis, $N_y$ the number of cells  packed along the $y$ axis, $N_z$ the number of cell layers, and $N_g$ the number of ghost node layers to cap the tissue. $N_x$ and $N_y$ give the spatial domain size as $\Omega_x = [0, N_x]$ and $\Omega_y = [0, \frac{\sqrt{3}}{2}N_y]$ to ensure hexagonal packing of cells.
Since we always require that an epithelial monolayer is present, the number of stromal cell layers is $N_z - 1$. For example, the tissue in Figure \ref{fig:initial_tissue} is of size $\left[ N_x, N_y, N_z, N_g\right] = \left[8,10,2,1\right]$.

\subsection{Cell turnover}
We choose proliferative cells to have a uniform cell cycle duration, $T \sim U(11,13)$, with a mean of 12 hours \cite{barker2009crypt,cheng1974origin}. When a proliferative cell completes its cell cycle, it is labelled a parent cell, with position $\mathbf{r}^p$. The parent cell proliferates, producing two daughter cells, with positions $\mathbf{r}_i$ and $\mathbf{r}_j$, separated by distance $\varepsilon >0$ apart. Following the division event, the parent cell is replaced by the two daughter cells placed $\varepsilon$ apart on the plane defined by the neighbours of the parent cell and the parent cell itself, spanned by the orthonorma vectors $\mathbf{\hat{u}}$ and $\mathbf{\hat{v}}$. We can write the daughter cell positions as:
\begin{align}
    \mathbf{r}_i = \mathbf{r}^p + \frac{\varepsilon}{2} \left( \cos\theta \mathbf{\hat{u}} + \sin \theta \mathbf{\hat{v}} \right), \quad \mathbf{r}_j = \mathbf{r}^p - \frac{\varepsilon}{2} \left( \cos \theta \mathbf{\hat{u}} + \sin \theta \mathbf{\hat{v}} \right),
\end{align}
where $\theta \in \left[0,2\pi\right)$ is drawn uniformly. We call this form of division \textit{planar cell division}. During a cell's first hour of cell cycle, we take the standard approach to increase the length separation between cell-centres $i$ and $j$ uniformly as:
\begin{align}
    s_{ij}(t) = \begin{cases}
     \varepsilon + \tau_i \left(1- \varepsilon\right), \quad &\tau_i \leq 1, \text{ and $i,j$ share a parent cell},\\
     1, \quad &\text{otherwise},
    \end{cases}
\end{align}
where $\tau_i \geq 0$ is the age of cell $i$ at time $t$ \cite{meineke2001cell,osborne2017comparing}.\\

We also know that if an epithelial cell detaches from the stroma, it is shed tissue and dies, a process known as anoikis \cite{yin2022missing,battini2006stable,eisenhoffer2012crowding}. To account for this, if an epithelial cell loses contact with the tissue, we remove it from the simulation.
Lastly, if cell $i$ is marked as being apoptotic, over the next hour, the cell shrinks by changing the length separation as:
\begin{align}
    s_{ij}(t) = \begin{cases}
     1 - 2 \tau^A_i, \quad & 0 \leq \tau^A_i \leq \frac{1}{2}\\
     0,  \quad & \frac{1}{2} < \tau^A_i \leq 1,\\
     1, \quad & \text{if $j$ is a stromal cell},
    \end{cases}
\end{align}
where $\tau^A_i$ is the time since apoptosis began for cell $i$.
\subsection{Implementation}
To simulate the tissue, we choose $\alpha = 1.01$ and $\beta = 4$ for the bending force parameters, as these provide the desired dynamics to maintain a flat, epithelial monolayer surface, see \ref{sec:model_cal}. We also choose a spring constant, $k_{i} = 20$ and $p_{ij}=1$ if $i$ and $j$ are the same cell type, otherwise $p_{ij} = 0.5$,
for the interaction model, and $\nu_i = 2$ for cell viscosity. We solve the equations of motion numerically using a Forward Euler method, with a step size of $\Delta t = 0.001$ hrs. Table \ref{tab:paramaters_values} contains a summary of the parameters used for the remainder of this paper. We implement the models used within this paper in the Chaste environment, which is an open source C++ library, available at \url{https://chaste.cs.ox.ac.uk} \cite{cooper2020chaste}. The code developed for this paper, is  freely available at \url{https://github.com/DGermano8/ModelingRealistic3DDeformationsOfEpitheliumInDynamicHomeostasis.git}.

\begin{table}[H]
\centering
\begin{tabular}{ c | c | c | c } 
 Parameter & Value & Units & Reference \\
 \hline
 $\alpha$ & 1.01 & - &  \ref{sec:model_cal} \\
 $\beta$ & 4 & CD$^{2}$ \, $\text{rad}^{-\alpha}$ &  \ref{sec:model_cal} \\
 $k_{i}$ & 20 & CM hrs$^{-2}$ & \cite{meineke2001cell}\\
 $p_{ij}$ & $\begin{cases}1, \, \text{if $i, \,j$ are same cell type}  \\0.5, \, \text{otherwise} \end{cases}$ & - & - \\
 $\nu_i$ & 2 & CM hrs$^{-1}$  & \cite{meineke2001cell} \\
 $\Delta t$ & 0.001 & hrs & - \\ 
\end{tabular}
\caption{\label{tab:paramaters_values} Parameters used for \textit{in silico} tissue simulations. Cell diameter (CD) is the natural spatial unit of the system and cell mass (CM) is the natural mass units of the system.}
\end{table}
An example of a tissue used initialise with size $\left[ N_x, N_y, N_z, N_g\right] = \left[8,10,2,1\right]$ is shown in Figure \ref{fig:initial_tissue}. 

To measure how flat a given tissue is, we calculate the mean point curvature, $\bar{G}$ as
\begin{align}
\bar{G} = \frac{1}{\left\vert N^E \right\vert} \sum_{j \in N^E} G_j.
\end{align}

\section{Results}
\label{sec:results}
We use our model to simulate three \textit{in silico} experiments: 
\begin{itemize}
    \item the relaxation of a deformable non-renewing differentiated tissue,
    \item a deformable renewing tissue, to demonstrate the  model's robustness while undergoing tissue renewal, in comparison to the base model,
    \item cell migration in deformable renewing tissue, to 
    demonstrate the model's robustness while undergoing tissue renewal, cell migration and cell removal, in comparison to a traditional fixed geometry model.
\end{itemize}

\subsection{Deformable non-renewing differentiated tissue}
\begin{figure}[H]
\centering

\begin{subfigure}[b]{0.48\textwidth}
\centering \text{}
\includegraphics[width = 0.6\textwidth]{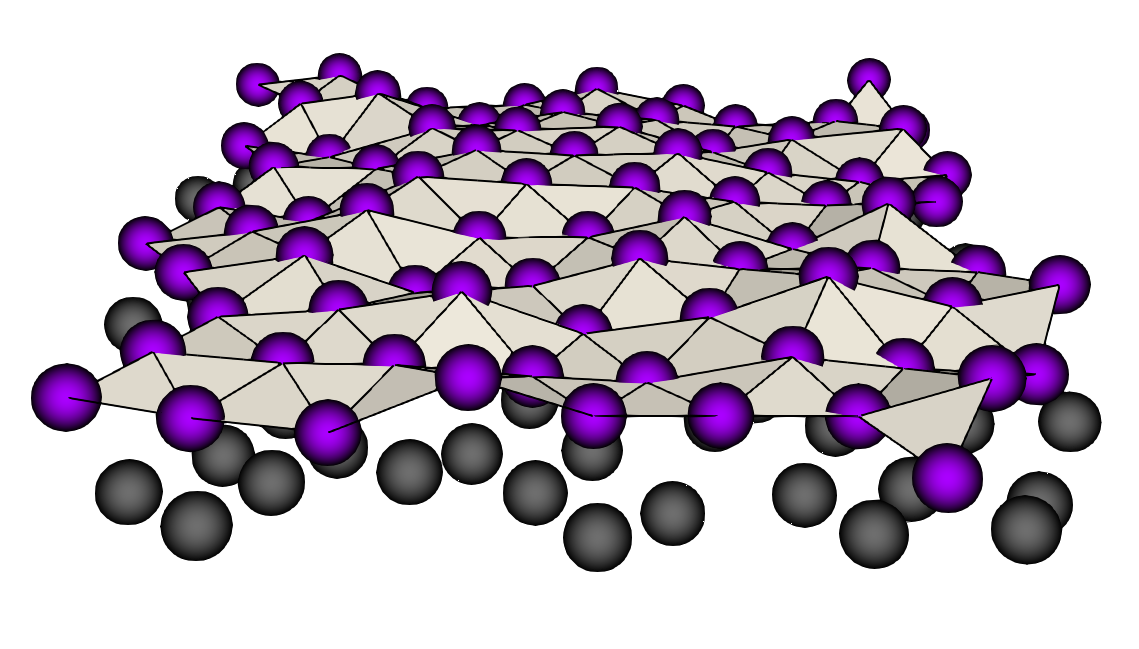}
\caption{\label{fig:tissue_sim_flat_0} $t=0$ hrs}
\end{subfigure}
\begin{subfigure}[b]{0.48\textwidth}
\centering \text{}
\includegraphics[width=0.6\textwidth]{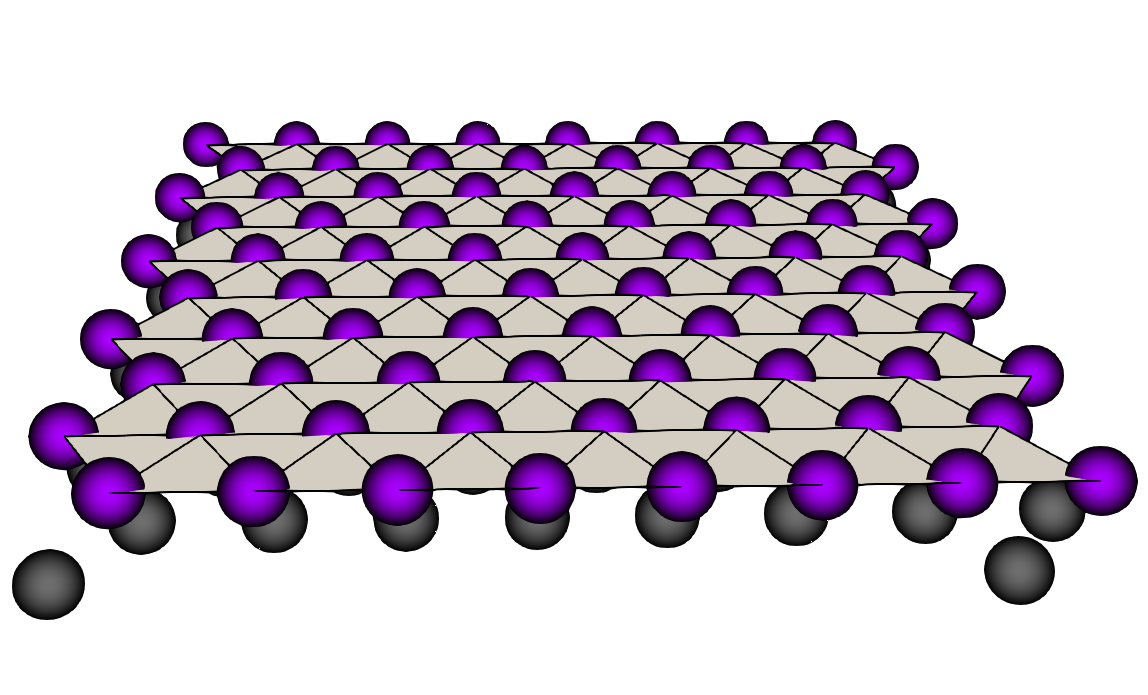}
\caption{\label{fig:tissue_sim_flat_2} $t=2$ hrs}
\end{subfigure}
\vspace{0.25cm}

\begin{subfigure}[b]{0.99\textwidth}
\centering \text{}
\includegraphics[width = 0.6\textwidth]{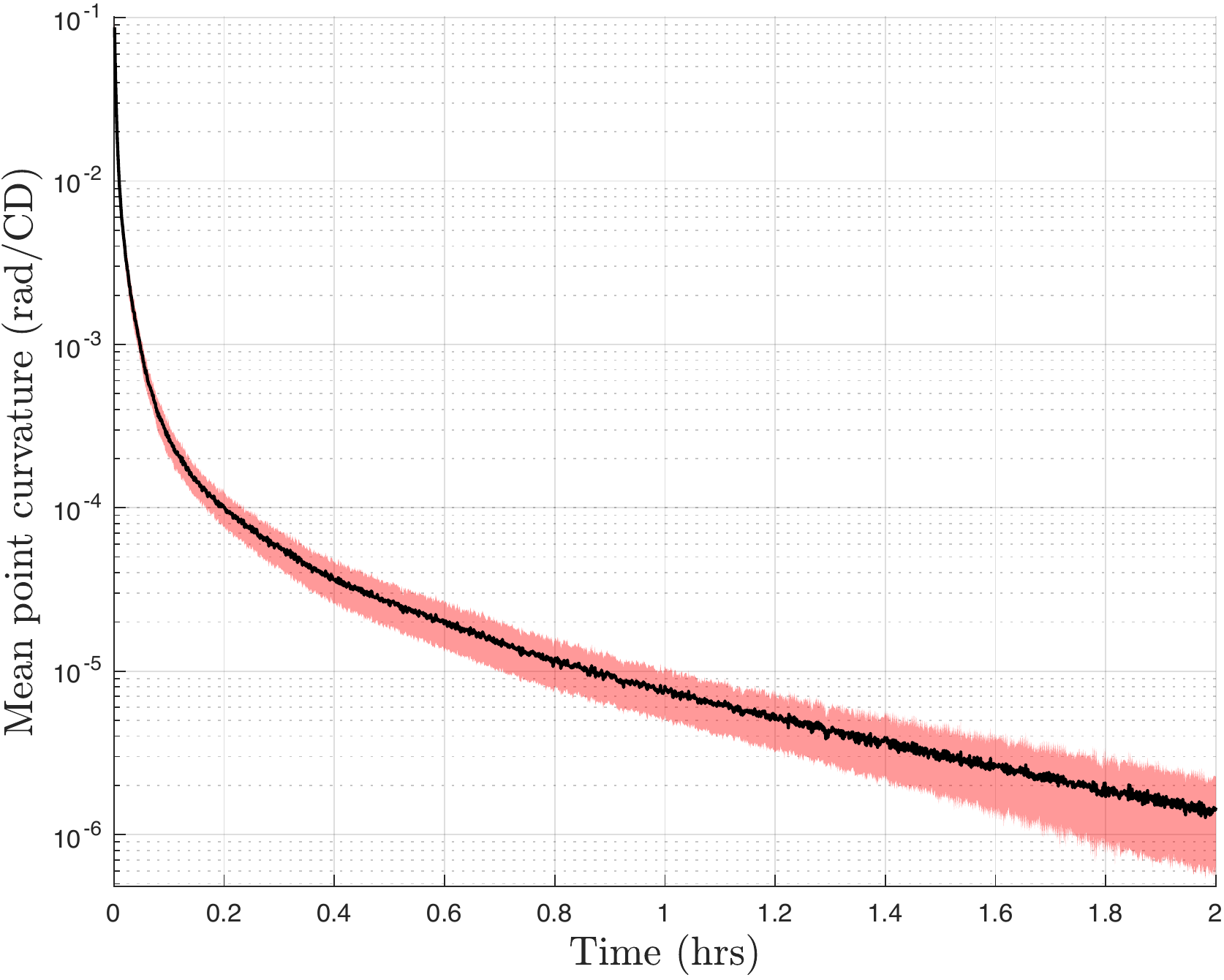}
\caption{\label{fig:curvature_with_time_flat} }
\end{subfigure}
\vspace{0.25cm}

\captionsetup{subrefformat=parens} 
\caption{ \label{fig:test_flatten} \textit{In silico} tissue simulations over 2 hours. Initial positions perturbed from equilibrium. We only show the differentiated epithelial cells (purple), the stromal cells (grey) and the epithelial monolayer surface $S^E$ shown via triangulation. \subref{fig:tissue_sim_flat_0} is sample tissue, perturbed from equilibrium, \subref{fig:tissue_sim_flat_2} is the final tissue. \subref{fig:curvature_with_time_flat} shows the mean point curvature over 20 sample tissue simulations (black line), with a 95\% confidence interval (red region). For an example video, see \ref{SI_Movie_1}}
\end{figure}

We first consider a tissue of differentiated cells relaxing. To do this, we take a tissue with hexagonal packing, at equilibrium, and perturb the $(x,y,z)$ positions by $(\zeta_x,\zeta_y,\zeta_z)$, where \hbox{$\zeta_i \sim U(-\frac{1}{4},\frac{1}{4})$}, for $i\in{x,y,z}$. An example of a perturbed tissue is shown in Figure \ref{fig:tissue_sim_flat_0}. The tissue is then free to relax, resulting in a  tissue similar to that shown in Figure \ref{fig:tissue_sim_flat_2}.  We take the average of 20 realisations of the \textit{in silico} tissue simulations, and calculate a 95\% confidence interval. We see from the mean point curvature in Figure \ref{fig:curvature_with_time_flat} that the curvature starts at the maximum, as the initial epithelial monolayer surface is not flat, and decreases as the bending force flattens the surface $S^E$. This demonstrates the model's ability to deform a non-flat tissue to flat over time.

\subsection{Deformable renewing tissue}
\begin{figure}[H]
\centering

\begin{subfigure}[b]{0.48\textwidth}
\centering \text{}
\includegraphics[width = 0.65\textwidth]{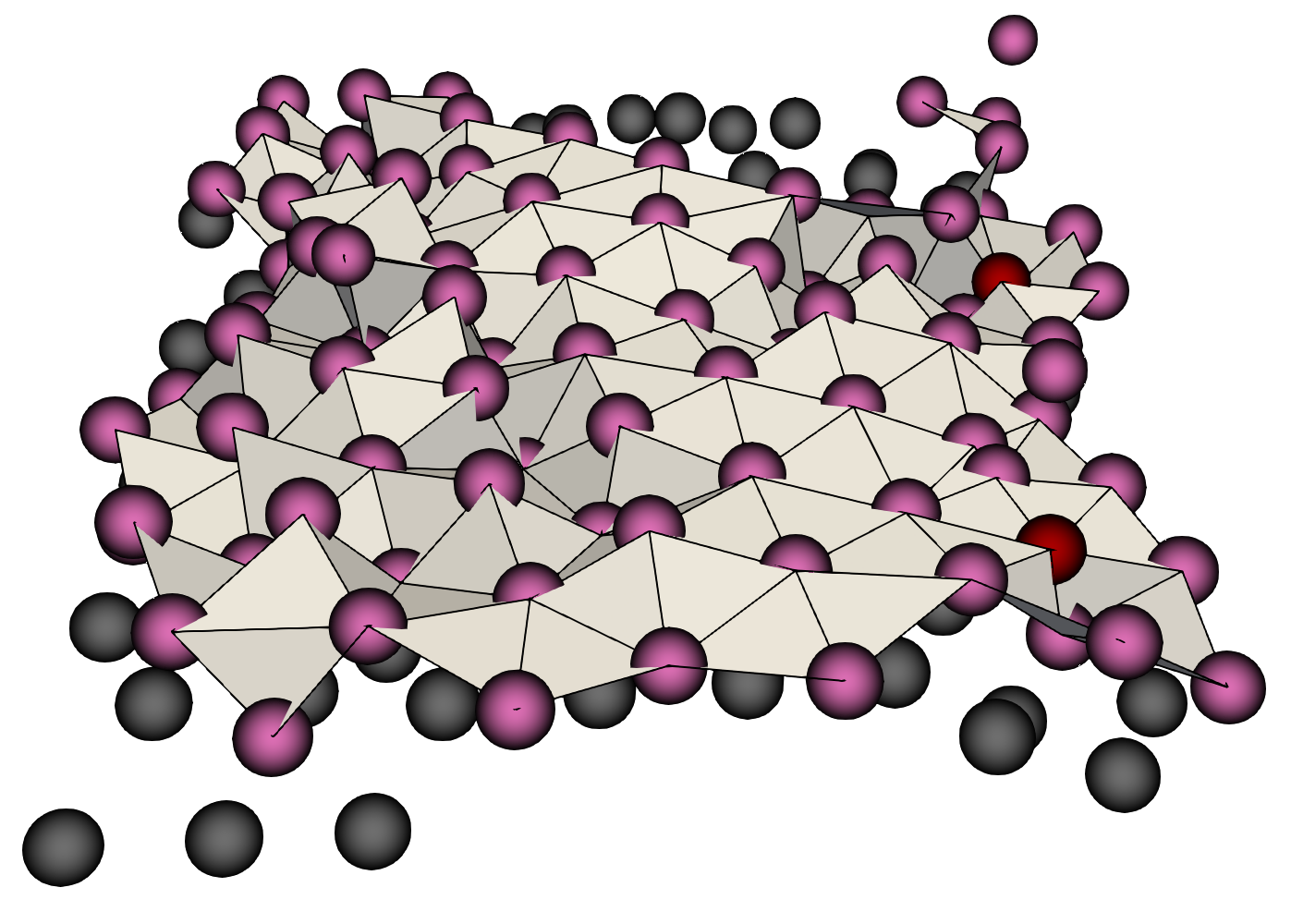}
\caption{\label{fig:NoPeriodicBend_RenewingTissue_48hrs_01_48hrs} Base model at $t=48$ hrs}
\end{subfigure}
\begin{subfigure}[b]{0.48\textwidth}
\centering \text{} 
\includegraphics[width=0.65\textwidth]{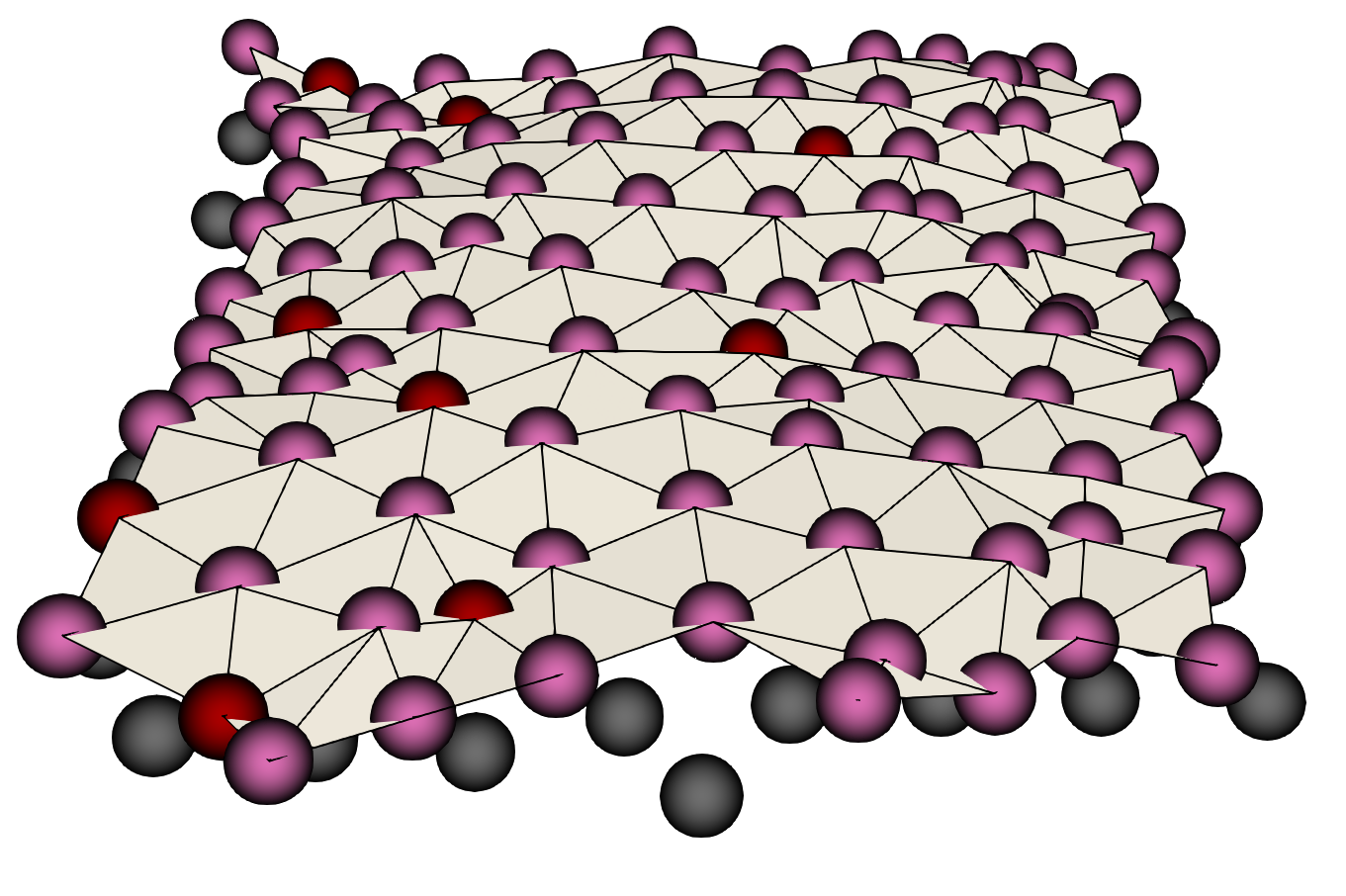}
\caption{\label{fig:PeriodicBend_RenewingTissue_48hrs_01_48hrs} Deformable model at $t=48$ hrs}
\end{subfigure}
\vspace{0.25cm}

\begin{subfigure}[b]{0.99\textwidth}
\centering \text{}
\includegraphics[width = 0.65\textwidth]{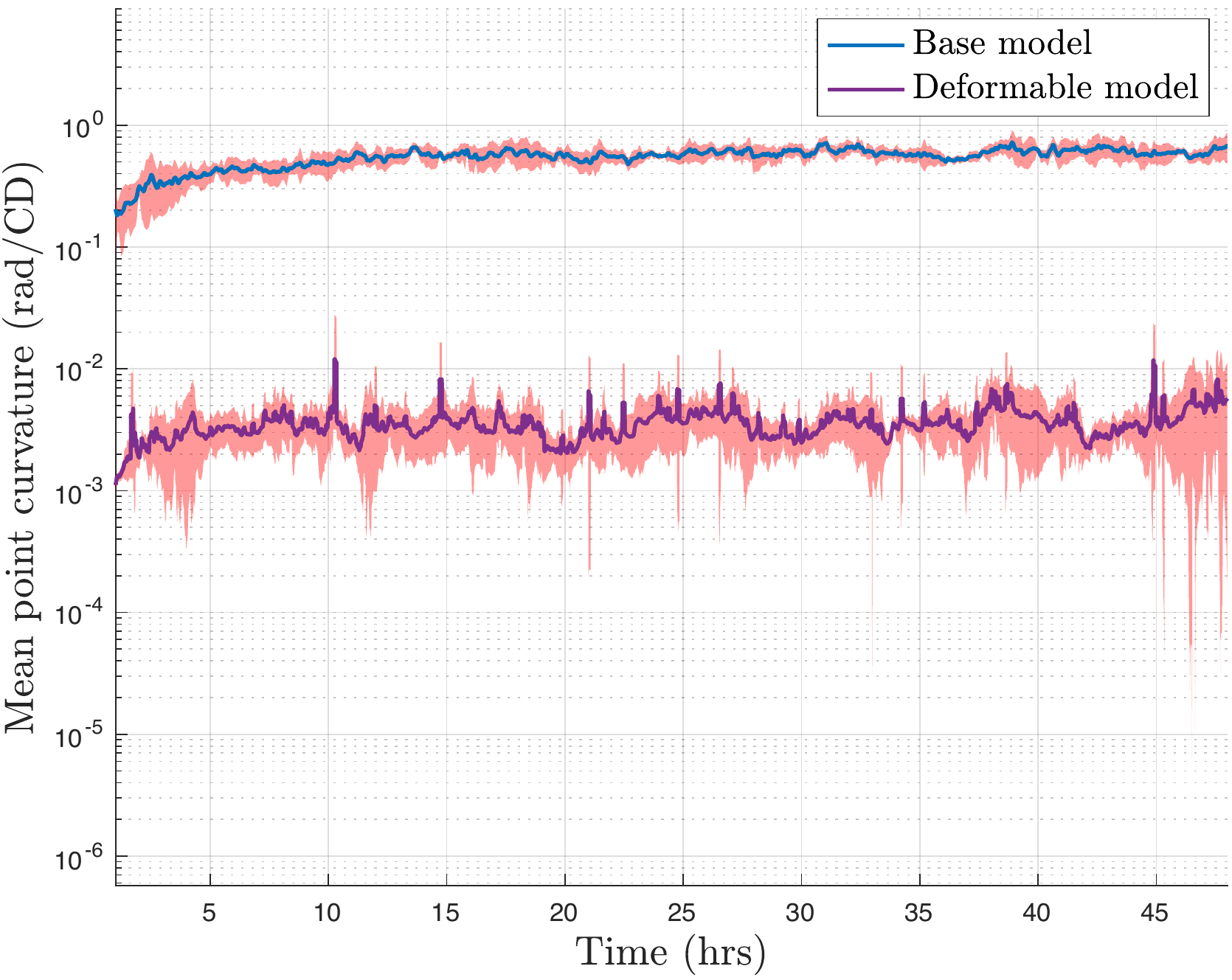}
\caption{\label{fig:curvature_with_time} }
\end{subfigure}
\vspace{0.25cm}

\captionsetup{subrefformat=parens} 
\caption{\label{fig:test_prol} Tissue simulations with proliferation and removal over 48 hours. For simplicity, we only show the epithelial cells, the epithelial monolayer surface $S^E$, and stromal cells.  \subref{fig:NoPeriodicBend_RenewingTissue_48hrs_01_48hrs} and \subref{fig:PeriodicBend_RenewingTissue_48hrs_01_48hrs} are the final states of the tissues for the base model and the deformable model respectively. 
\subref{fig:curvature_with_time} shows how the mean point curvature of the tissues varies with time, as cells proliferate. Blue shows the tissue for the base model, and purple the tissue with the deformable model, along with 95\% confidence intervals (red regions, calculated over 5 unique simulations). For an example video, see \ref{SI_Movie_2}.}
\end{figure}

We next consider a renewing tissue, with cell proliferation and removal, with results shown in Figure \ref{fig:test_prol}. To maintain a uniform tissue density and prevent overcrowding, when a cell proliferates, we mark another epithelial cell as apoptotic, chosen at random, unless an anoikis event has just occurred.

First, we simulate the tissue with no bending force, which we will call the \textit{base model}, and then again with identical initial conditions, this time with the bending force, which we will call the \textit{deformable model}, with examples of the final state of the tissue model shown in Figures \ref{fig:NoPeriodicBend_RenewingTissue_48hrs_01_48hrs} and \ref{fig:PeriodicBend_RenewingTissue_48hrs_01_48hrs}. We compute $\bar{G}$ for both the \textit{base model} and the \textit{deformable model}. 
We show how the mean point curvature varies as cells proliferate in Figure \ref{fig:curvature_with_time}.
We  see that the base model develops kinks and is no longer flat, whereas with the deformable model, the tissue remains flat. We can also see that the mean point curvature for the two different tissues varies by up to 3 orders of magnitude, with the deformable model having a significantly lower mean point curvature. Lastly, we note that the deformable model contains distinct spikes in the mean point curvature, which are not sustained, but disappear as the tissue flattens. These spikes are caused by cells proliferating. We note that the same spikes are also present in the base model, but appear less significant due to the log scale. These results show that the deformable models ability to maintain structure is robust while the tissue undergoes self renewal.

\subsection{Cell migration in deformable renewing tissue}
\begin{figure}[H]
\centering
\includegraphics[width = 0.7\textwidth]{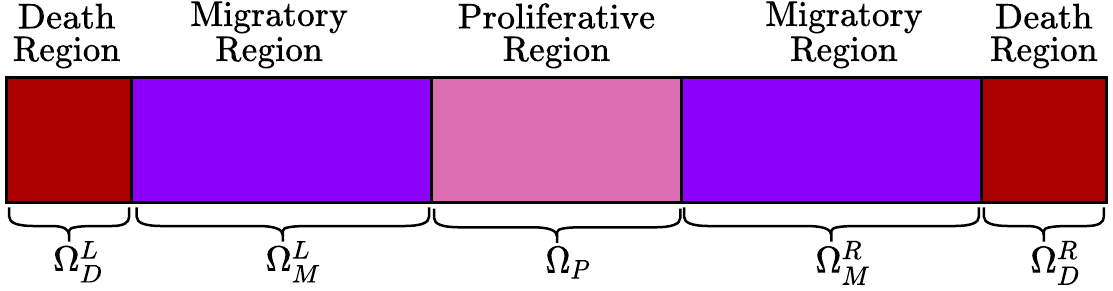}
\caption{\label{fig:WntFlowRegion} Domain structure with the proliferative, $\Omega_P$, migratory, $\Omega_M =\Omega^L_M \cup \Omega^R_M$, and death, $\Omega_D = \Omega^L_D \cup \Omega^R_D$, regions.}
\end{figure}

We now consider both proliferative epithelial cells and differentiated epithelial cells, to observe how our model behaves with cell migration.  We implement a simple position based model, where a cell's position determines its cell type. In the proliferative region, with $\mathbf{r}_{i} \in \Omega_P$, the cell experiences high external signalling factor and the cell is proliferative, otherwise a cell is differentiated. 
We also implement a density dependent death model where, if a cell's area contribution is below a certain threshold (i.e. $A_j < A_{\text{crit}}$), and the cell is located within the death region $\mathbf{r}_{i} \in \Omega_D = \Omega^L_D \cup \Omega^R_D$, the cell is marked as apoptotic and undergoes apoptosis over the next hour. Figure \ref{fig:WntFlowRegion} shows a diagram of the proliferative, migratory and death regions.

\begin{table}[H]
\centering
\begin{tabular}{ c | c | c  } 
 Parameter & Value & Units \\
 \hline
 $A_{\text{crit}}$ & $0.98 \, A_{\text{Relaxed}}$ & CD$^2$  \\
 $A_{\text{Relaxed}}$ & $\frac{\sqrt{3}}{2}$ & CD$^2$  \\
 $\Omega_P$ & $\{ (x,y,z)\in \mathbb{R}^3 : x\in[-1.5,1.5] \}$ & (CD,CD,CD) \\
 $\Omega_M$ & $\{ (x,y,z)\in \mathbb{R}^3 : x\in[-5.5,-1.5) \cup (1.5,5.5] \}$ & (CD,CD,CD) \\
 $\Omega_D$ & $\{ (x,y,z)\in \mathbb{R}^3 : x\in[-6,-5.5) \cup (5.5,6] \}$ & (CD,CD,CD) \\
\end{tabular}
\caption{\label{tab:Wnt_paramaters_values} Parameters used for \textit{in silico} tissue simulations. Cell diameter (CD) is the natural spatial unit of the system. $A_{\text{Relaxed}}$ is the area of a hexagon, which is the resting equilibrium of cells in hexagonal close packing.}
\end{table}

Using the domain structure shown in Figure \ref{fig:WntFlowRegion}, and the domain sizes and critical area for density-dependent death given in Table \ref{tab:Wnt_paramaters_values}, we simulate a tissue of size $\left[ N_x, N_y, N_z, N_g\right] = \left[12,14,3,1\right]$ for 480 hours.

\begin{figure}[H]
\centering

\begin{subfigure}[b]{0.99\textwidth}
\centering
\includegraphics[width = 0.8 \textwidth]{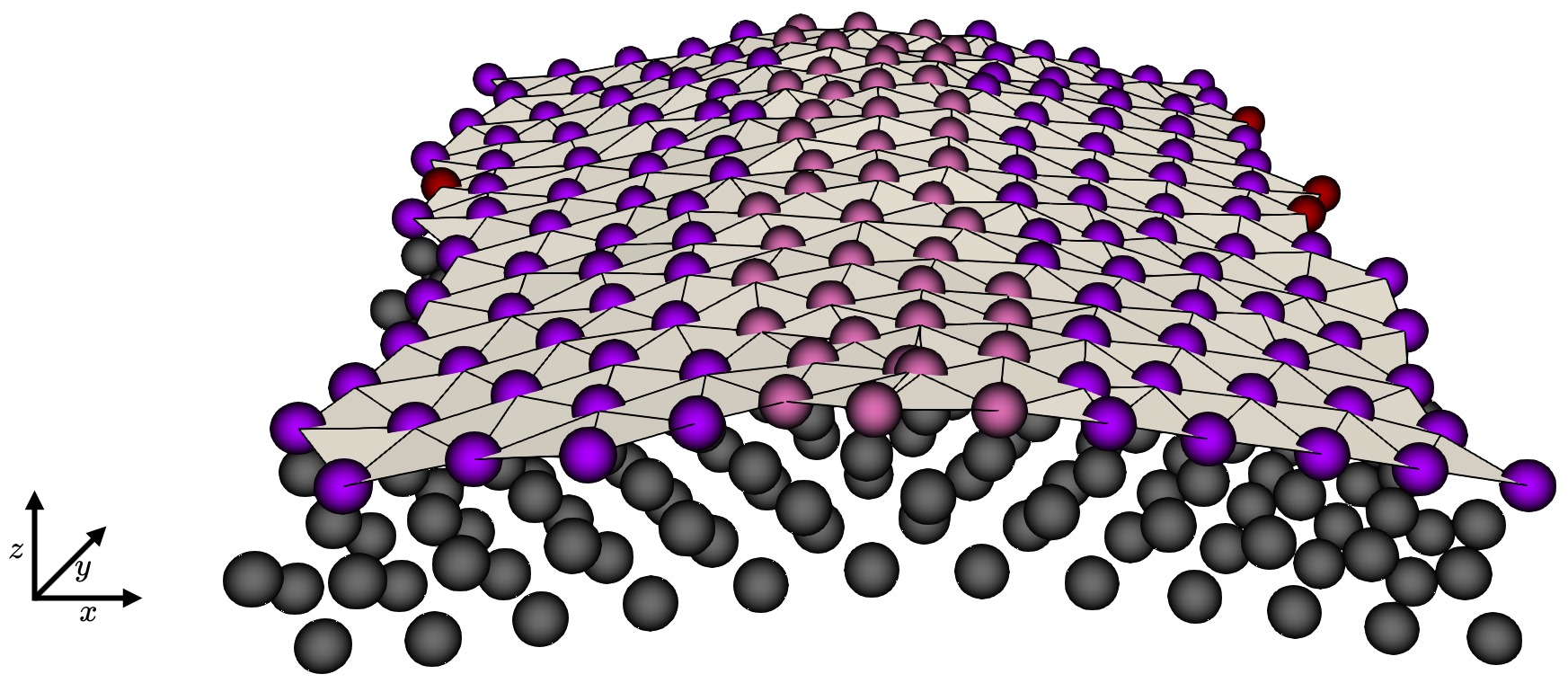}
\caption{\label{fig:WntFlow_42000}}
\end{subfigure}

\begin{subfigure}[b]{0.475\textwidth}
\centering \text{}
\includegraphics[width = 0.32\textheight]{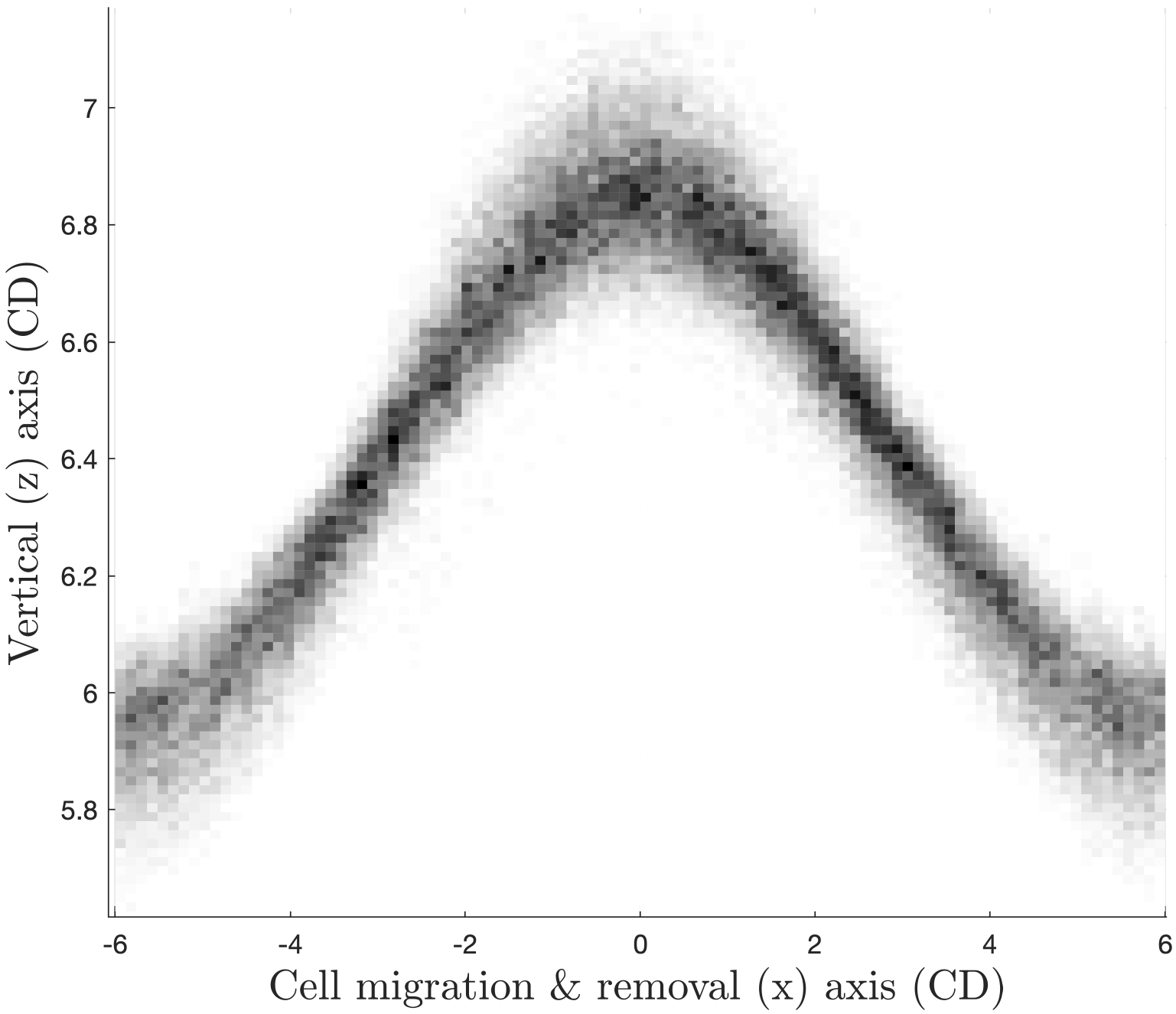}
\caption{\label{fig:ZX_Distribution}}
\end{subfigure}
\begin{subfigure}[b]{0.475\textwidth}
\centering \text{} 
\includegraphics[width=0.32\textheight]{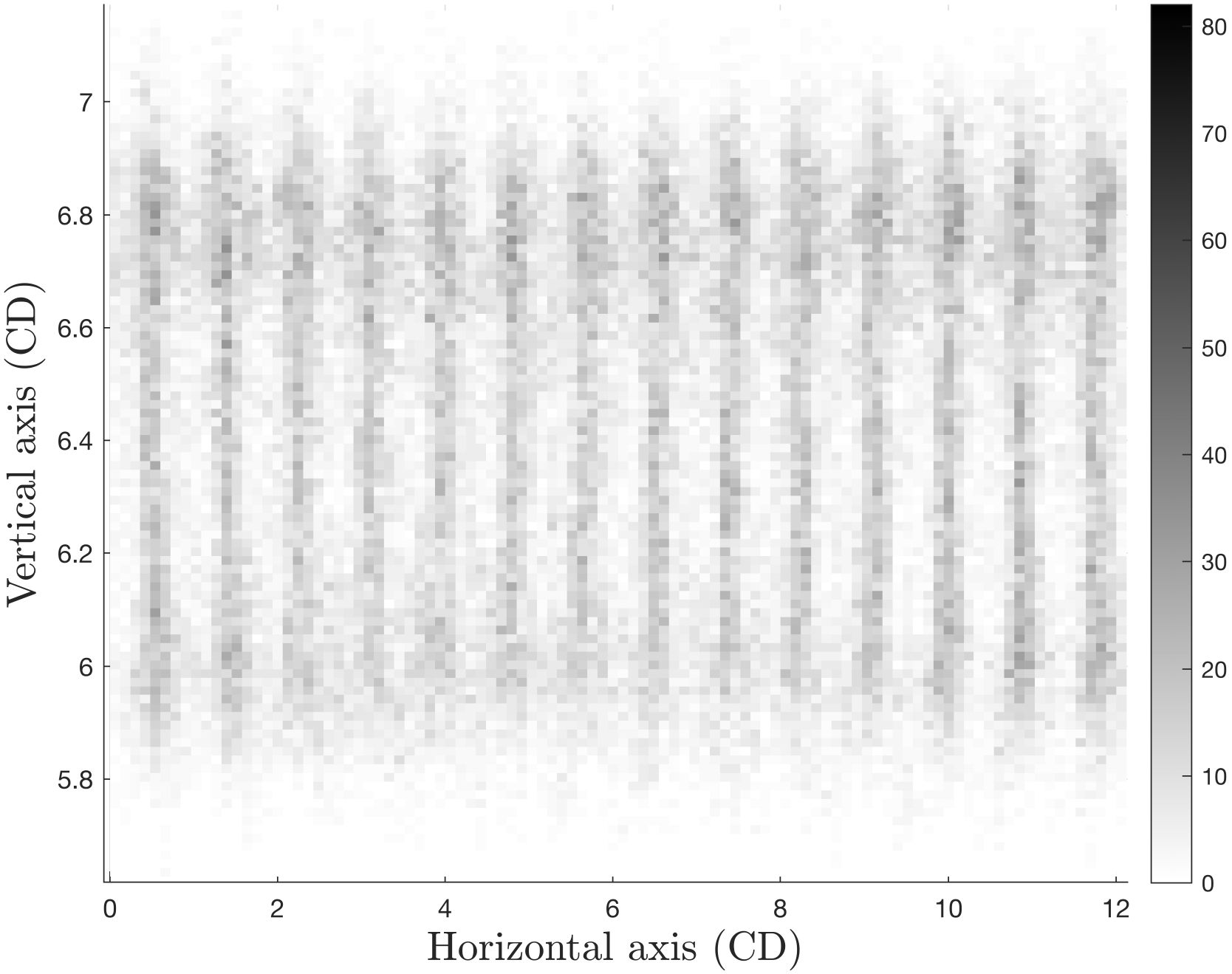}
\caption{\label{fig:ZY_Distribution} }
\end{subfigure}
\begin{subfigure}[b]{0.03\textwidth}
\rotatebox{90}{\scriptsize\text{Count}}
\vspace{3.5cm}
\end{subfigure}

\captionsetup{subrefformat=parens} 
\caption{An example of the tissue simulation at steady state. Figure \subref{fig:WntFlow_42000} shows hyper stimulated proliferative transient cells, shown in pink, migratory differentiated cells, shown in purple, and dying cells, shown in red. Stromal cells are shown in grey, and ghost nodes are not shown. Figures \subref{fig:ZX_Distribution} and \subref{fig:ZY_Distribution} show epithelial cell distributions throughout the tissue at homeostasis. The sample distributions are collated by sampling the tissue every hour for the last 360 hours of the simulation to ensure homeostasis. For an example video, see \ref{SI_Movie_3}}
\end{figure}

To ensure the tissue is in dynamic homeostasis, we discard the first 120 hours, and consider the tissue from time $t=120hrs$ to $t=480hrs$. In Figures \ref{fig:ZX_Distribution} and \ref{fig:ZY_Distribution} we show the cell distribution for the tissue in homeostasis, from 360 hourly samples of a simulation. In Figure \ref{fig:ZX_Distribution} we show that the tissue undergoes deformation in the vertical ($z$) axis with respect to the migration and removal ($x$) axis. However, this deformation is small at approximately 1CD across the 12CD migration and removal ($x$) axis. We can also see that, with respect to the migration and removal ($x$) axis, cells are uniformly distributed throughout. If we consider Figure \ref{fig:ZY_Distribution} which shows the cell distribution with the horizontal ($y$) and the vertical ($z$) axis. Here we can see that cells occupy distinct strips within the tissue, which correspond to the migration paths of the cells through the tissue. We note that these distinct bands are a result of the domain specification chosen.


To observe how the model presented here behaves in comparison to previous models of cell migration, we consider a tissue evolving on a fixed geometry and compared it to the above simulation. The fixed geometry is such that cells are free to move along the migration and removal axis, as well as the horizontal axis, but epithelial cells are limited to a fixed vertical height, $z_F$. Stromal cells and  ghost nodes do not have any restrictions. We note that this is similar to considering a purely 2D model of cell migration. However, a 2D model is not sufficient as it does not account for the cell-cell interactions between epithelial and stromal cells, which has the effect of an increased effective viscosity in 2D. As such, we compare our model to a fixed geometry model rather than a purely 2D model.

\begin{figure}[H]
\centering

\begin{subfigure}[b]{0.49\textwidth}
\centering \text{}
\includegraphics[width = 0.99\textwidth]{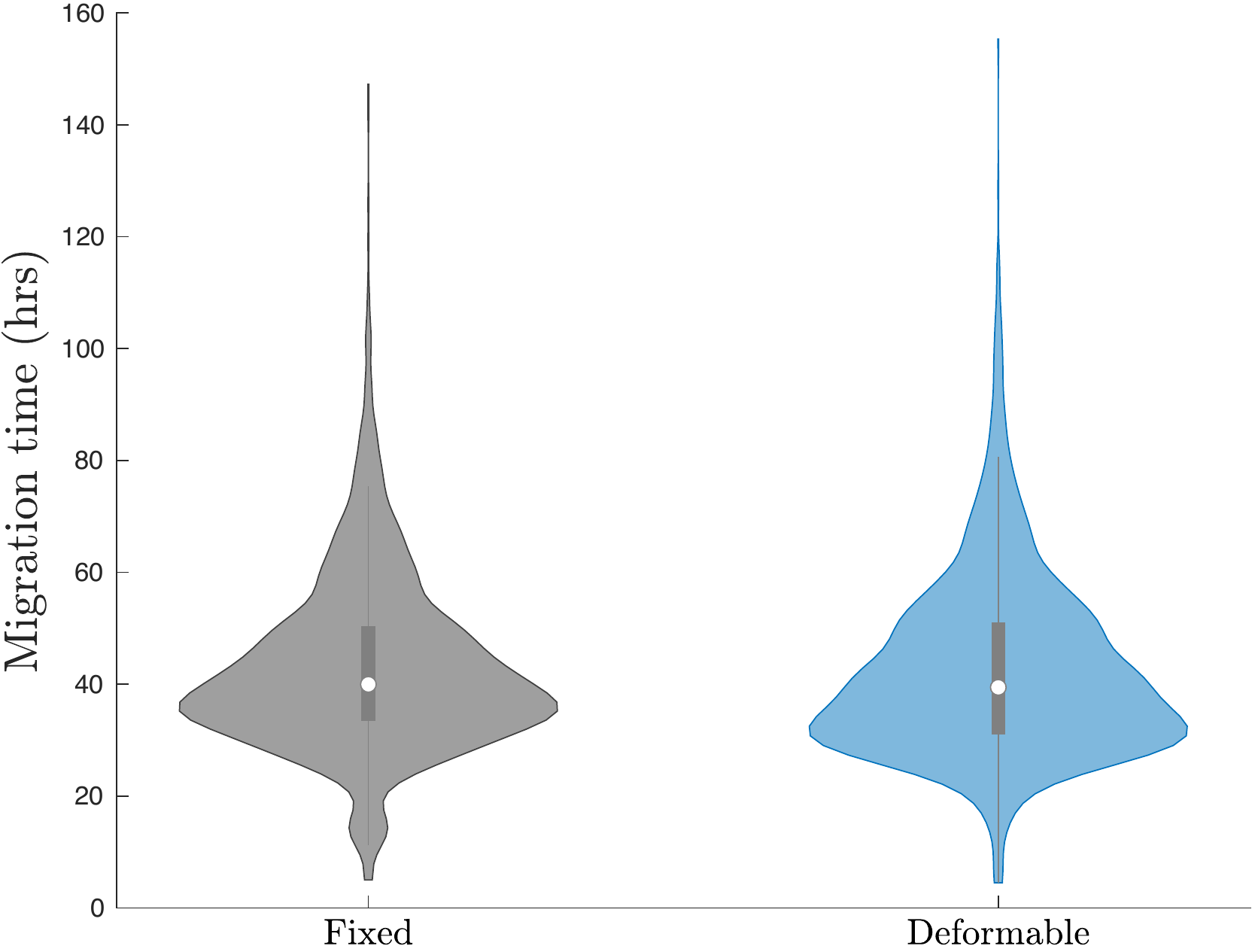}
\caption{\label{fig:migrationTimes}}
\end{subfigure}
\begin{subfigure}[b]{0.49\textwidth}
\centering \text{} 
\includegraphics[width=0.99\textwidth]{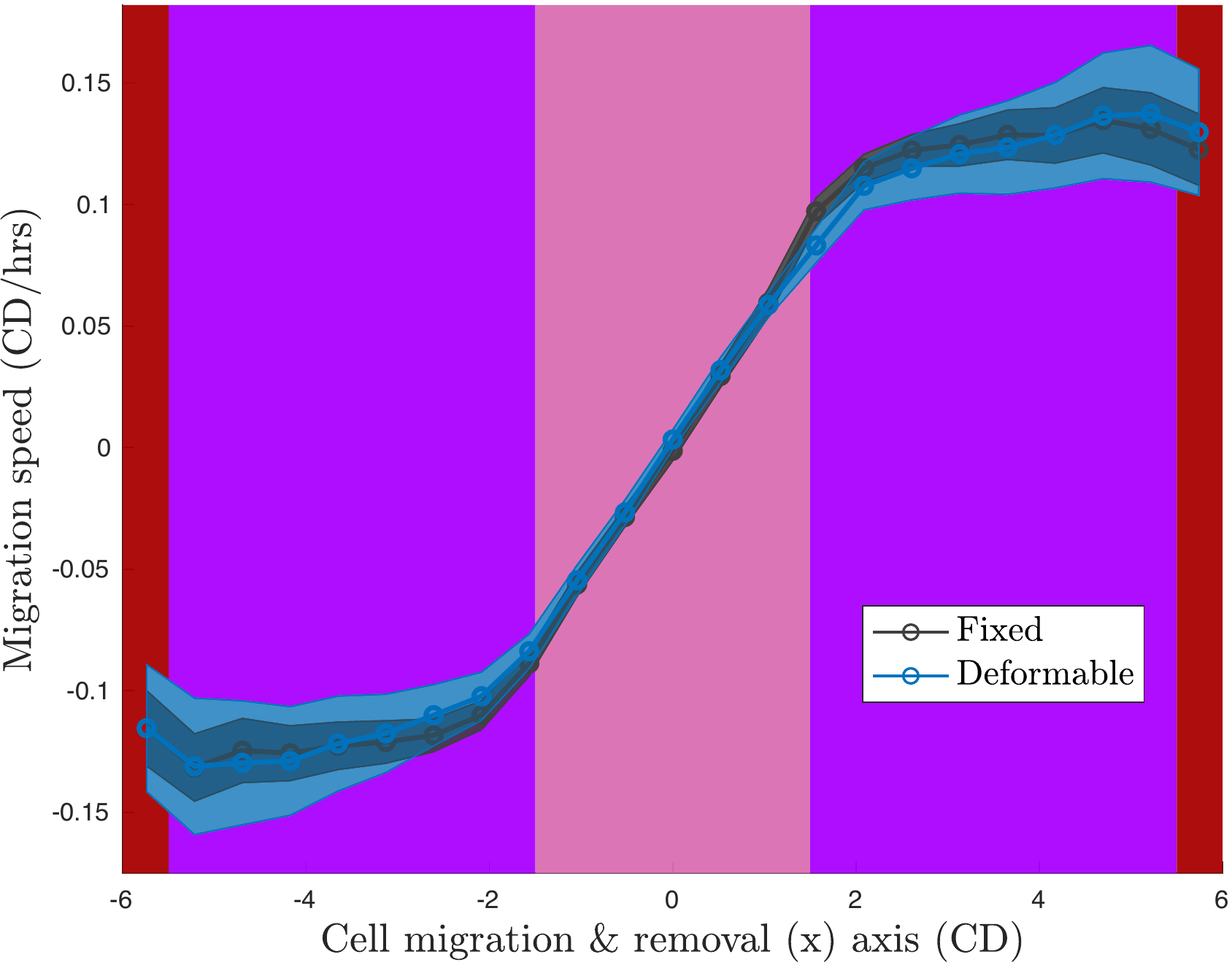}
\caption{\label{fig:MigrationSpeed} }
\end{subfigure}

\caption{\label{fig:ComparisionTissues}Distribution of migration times for fixed (grey) and deformable (blue) models are shown in \ref{fig:migrationTimes}. The migration speeds along the cell migration and removal ($x$) axis are shown in \ref{fig:MigrationSpeed}, for both the fixed (grey) and deformable (blue) models, each with 95\% confidence intervals. The background colours of Figure \ref{fig:MigrationSpeed} show  the region the sample point is in.}
\end{figure}

Figure \ref{fig:migrationTimes} shows the distribution of migration times for both the fixed and deformable models. The migration time is defined as the time taken for a cell to migrate throughout the migratory region, $\Omega_M$. We see that while the deformable model has a longer tail, and is slightly skewed to lower migration time, both tissues behave in a similar manner.

If we discretize the migration and removal axis, and look at a cells migration speed within each region, we observe how cells are migrating throughout the tissue, as shown in Figure \ref{fig:MigrationSpeed}, for both fixed (grey) and deformable (blue) models. We can see that, within the proliferative region, $\Omega_P$, cell migration speeds overlap. At the interface between the proliferative and migratory region, the fixed model has a higher cell migration speed. However throughout the remainder of the tissue, migrations speeds agree, within the margin of error of the 95\% confidence intervals.

The analysis shown in Figure \ref{fig:ComparisionTissues} indicates that the deformable model does not unduly influence the dynamics found in fixed geometry models of dynamic homeostasis, while still allowing deformations to be included in such systems. These results show that the deformable models ability to maintain structure is robust while the tissue undergoes renewal, cell migration and cell removal.

\section{Discussion}
\label{sec:conclusion}
We have presented a novel model of a 3D deformable, dynamic tissue. Using a cell-centre based Delaunay triangulation model, we described cell interactions via a spring potential, and using the same triangulation, we developed a bending potential to describe the shape of a tissue. Through the minimisation of this bending  potential, we displayed how we are able to maintain a flat epithelial monolayer, undergoing proliferation and death, which resides on a stromal tissue.

We then compared how a simulated tissue behaves both with and without a bending force acting on the tissue. We found that when we do not include a bending force, the tissue develops kinks and no longer remains flat. In comparison, when we included the bending force, we were able to maintain a flat epithelial monolayer, with a bending potential of up to 3 orders of magnitude smaller than the tissue without a bending force.

Finally, we implemented the model on a tissue which experiences localised cell proliferation, and a local density-dependent death model. We found that with our model presented here, we are able to achieve cell migration while still maintaining a relatively flat epithelial monolayer. We also compared our model to a conventional 3D fixed geometry model, and found that the migration time and migration speed follow similar distributions. From this, we can conclude that the model presented here is well suited to capture the compartmental and deformable nature of a realistic tissue in dynamic homeostasis.

Future avenues of study include incorporating a realistic curved tissues, and to determine the conditions under which does the system remains stable. We will then apply the model to some commonly studied tissues and organs. We would also like to compare the results here with those obtained from continuum models of tissue deformation and cell-based models of deformable tissues, to draw similarities and differences.

\section*{Acknowledgements}

This research was supported by an Australian Government Research Training Program (RTP) Scholarship (awarded to DPJG). S.T.J. is supported by the Australian Research Council (project no. DE200100998).

\bibliography{bibModelingRealistic3DDeformationsOfEpitheliumInDynamicHomeostasis}




\newpage
\renewcommand{\thepage}{SI.\arabic{page}}
\setcounter{page}{1}
\renewcommand{\thesection}{SI.\arabic{section}}
\setcounter{section}{0}
\renewcommand{\thetable}{SI.\arabic{table}}
\setcounter{table}{0}
\renewcommand{\thefigure}{SI.\arabic{figure}} 
\setcounter{figure}{0}
\renewcommand{\theequation}{SI.\arabic{equation}}
\setcounter{equation}{0}

\section*{Supplementary Information}
\section{Equations of motion}
 \label{sec:eq_of_motion}
To calculate the exact form of the bending potential, we take the gradient with respect to cell $i$ and note that since only the triangular components of $\phi_j$ that contain cell centre $i$ will contribute:
\begin{align}
    \mathbf{F}^{\text{Bending}}_{i} &= - \nabla_i \sum_{j \in N^E} \beta_j \left\vert \frac{ 2\pi - \phi_j }{A_j} \right\vert ^{\alpha}, &&\forall i \in N^E,\\
    &= - \sum_{j \in N^\text{E}_i} \beta_{i} \nabla_i  \left\vert \frac{ 2\pi - \phi_j }{A_j} \right\vert ^{\alpha}, &&\forall i \in N^E,\\
    &=  - \alpha \sum_{j \in N^\text{E}_i} \beta_{i}  \text{sgn}\left( \frac{ 2\pi - \phi_j }{A_j} \right) \left\vert \frac{ 2\pi - \phi_j }{ A_j } \right\vert ^{\alpha-1} \, \left[ \frac{  A_j \nabla_i \phi_j - (2 \pi - \phi_j ) \nabla_i A_j }{A_j^2}  \right], &&\forall i \in N^E,
\end{align}
where $N^\text{E}_i$ are the first-order, epithelial neighbours of cell-centre $i$. From here, we see how $\beta_{i}$ describes the magnitude of the bending force, and $\alpha$ the sensitivity of the bending force, which are further discussed in \ref{sec:model_cal}. To maintain our epithelial monolayer surface, $S^E$, the bending force is only applied to the epithelial cells. The gradient of $\phi_j$, with respects to cell $i$ is:
\begin{align}
    \mathbf{\nabla}_i \phi_j = 
    \begin{cases}
     \frac{-1}{\sqrt{1-\cos^2(\theta_{mjn})}}  \left[ \left(\frac{\cos(\theta_{mjn})}{|\mathbf{r}_{jm}|} - \frac{1}{|\mathbf{r}_{jn}|} \right)\mathbf{\hat{r}}_{jm} + \left(\frac{\cos(\theta_{mjn})}{|\mathbf{r}_{jn}|} - \frac{1}{|\mathbf{r}_{jm}|} \right)\mathbf{\hat{r}}_{jn}  \right], \quad &\text{if } j=i,\\
     \frac{-1}{\sqrt{1-\cos^2(\theta_{jim})}} \frac{1}{|\mathbf{r}_{ij}|} \left[ \mathbf{\hat{r}}_{im} - \mathbf{\hat{r}}_{ij} \cos(\theta_{jim}) \right], \quad &\text{if } j \in N^E_i,\\
     \mathbf{0}, \quad &\text{otherwise},
    \end{cases}
\end{align}
where $\theta_{mjn}$ is the angle between $\mathbf{r}_{jm}$ and $\mathbf{r}_{jn}$. Similarly, the gradient of $A_j$, with respect to cell $i$ is:
\begin{align}
    \mathbf{\nabla}_i A_j = 
    \begin{cases}
    \frac{1}{4A_j}\left[ \psi_n\left(\mathbf{r}_{nm},\mathbf{r}_{mj}\right) +  \psi_m\left(\mathbf{r}_{mn},\mathbf{r}_{nj}\right) +   \psi_j\left(\mathbf{r}_{mn},\mathbf{r}_{mn}\right) \right]  , \quad &\text{if } j \in N^E_i,\\
     \mathbf{0}, \quad &\text{otherwise},
    \end{cases}
\end{align}
where $\psi_l\left(\mathbf{r}_{ab},\mathbf{r}_{cd}\right) = \mathbf{r}_{l} \left( \mathbf{r}_{ab} \cdot \mathbf{r}_{cd} \right) - \mathbf{r}_{l} \cdot \left( \mathbf{r}_{ab} \circ \mathbf{r}_{cd} \right)$, and $  \mathbf{r}_{ab} \circ \mathbf{r}_{cd} $ is the Hadamard product between vectors $ \mathbf{r}_{ab}$ and $ \mathbf{r}_{cd} $.
From here, we can rearrange to give the following equations of motion:
\begin{align}
    \nu_i \frac{d \mathbf{r}_i}{dt} = \sum_{j \in N_i} k_{i} p_{ij} \left( \vert \mathbf{r}_{ij} \vert - s_{ij} \right) \mathbf{\hat{r}}_{ij}  - \sum_{j \in N^\text{E}_i} \alpha \beta_{i}  \text{sgn}\left( \frac{ 2\pi - \phi_j }{A_j} \right) \left\vert \frac{ 2\pi - \phi_j }{ A_j } \right\vert ^{\alpha-1} \, \left[ \frac{  A_j \nabla_i \phi_j - (2 \pi - \phi_j ) \nabla_i A_j }{A_j^2}  \right], \quad \forall i.
\end{align}
We now define the magnitude of the bending potential at cell $j$, $\beta_j$, in terms of the spring constant $k_j$, and a proportionality constant, $\rho_j$, which shares the same non-zero properties as $\beta_j$, and write:
\begin{align}
    \Rightarrow  \nu_i \frac{d \mathbf{r}_i}{dt} &= \sum_{j \in N_i} k_{i} \left\{ p_{ij} \left( \vert \mathbf{r}_{ij} \vert - s_{ij} \right) \mathbf{\hat{r}}_{ij}  - \alpha \rho_{i} \,  \text{sgn}\left( \frac{ 2\pi - \phi_j }{A_j} \right) \left\vert \frac{ 2\pi - \phi_j }{ A_j} \right\vert ^{\alpha-1} \, \left[ \frac{  A_j \nabla_i \phi_j - (2 \pi - \phi_j ) \nabla_i A_j }{A_j^2}  \right]  \right\}, \quad \forall i.
\end{align}

\section{Bending force calibration}
 \label{sec:model_cal}
\begin{figure}[H]
\centering
\includegraphics[width = 0.55\textwidth]{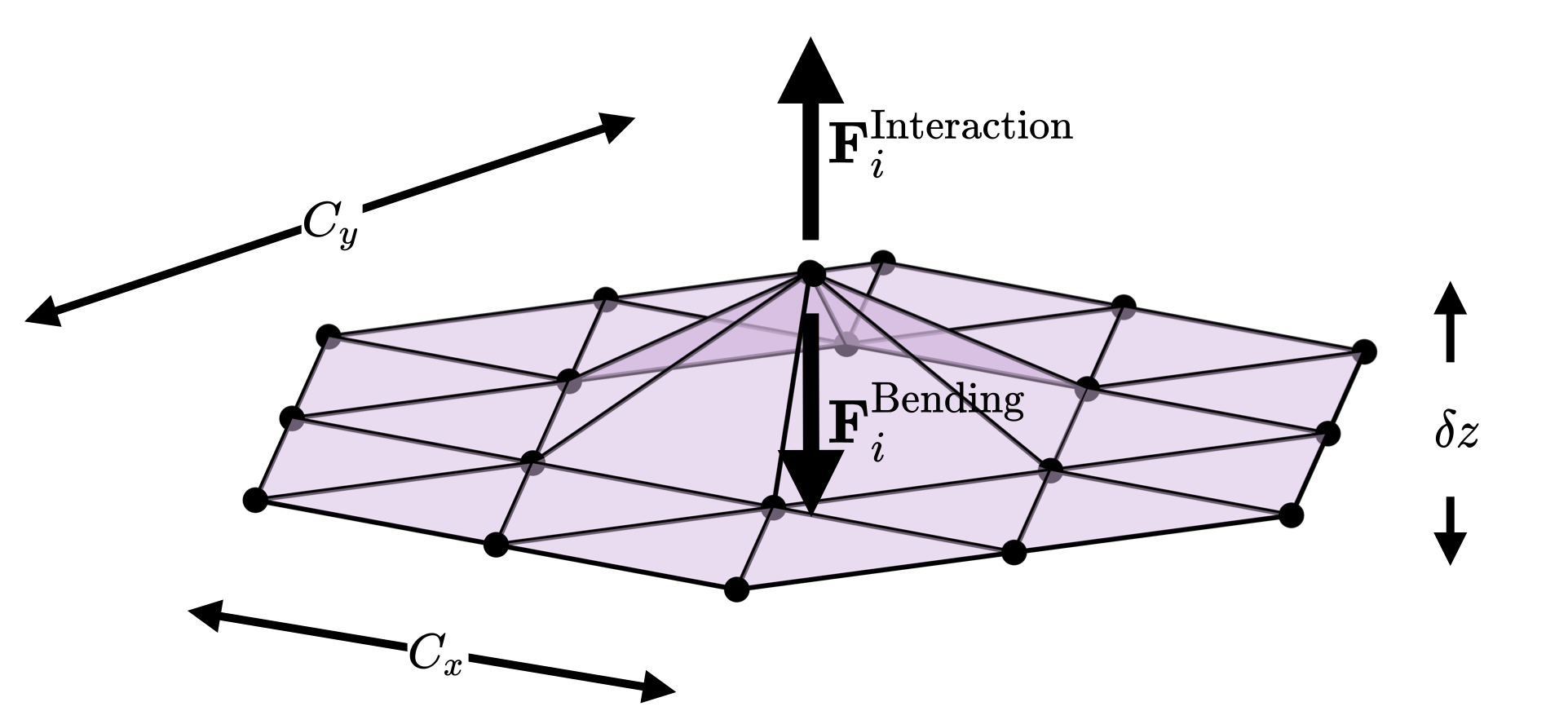}
\caption{\label{fig:config_elem} Test element used to calibrate model. $C_x$ and $C_y$ are the amount of compression in the $x$ and $y$ components respectively, and  $\delta z$ the $z$ displacement from equilibrium.}
\end{figure}

To calibrate the bending force in our model, we consider a test element at equilibrium, and perturb the $z$-position of the centre cell by $\delta z$. We also consider the effect compression has on the model, where $C_x$ and $C_y$ are the axis scalings in the $x$ and $y$ components respectively, and where $C_x = C_y = 1$ is no compression. The test element is shown in Figure \ref{fig:config_elem}. We see that since planar forces in $\mathbf{F}_i^{\text{Interaction}}$ and $\mathbf{F}_i^{\text{Bending}}$ each individually balance, the contribution due to these forces will be in the $z$-component only. Specifically $\mathbf{F}_i^{\text{Bending}}$ will act downwards, and $\mathbf{F}_i^{\text{Interaction}}$ will act upwards. We therefore only need to consider the magnitudes of $\mathbf{F}_i^{\text{Bending}}$ and   $\mathbf{F}_i^{\text{Interaction}}$ .

Before we perform our calibration, we scale the bending force with respect to the interaction force, writing $\beta_i$ in terms of $k_i$ and a proportionality variable, $\rho_i$, which shares the same non-zero properties as $\beta_i$, and define $\beta = k_i \rho$. We can then rewrite Equation (\ref{eq:motion_eq_1}) as:
\begin{align}
\nu_i \frac{d \mathbf{r}_i}{dt} = -\nabla_i  \sum_{\forall j} k_j \left( P_j +   \rho_j \vert G_j \vert^\alpha \right), \quad \forall i.
\end{align}

We now use the test element in Figure  \ref{fig:config_elem} to determine the exponent parameter, $\alpha$, and the proportionality constant, $\rho$. Since we will have equal compression in the $x$ and $y$ axes, we denote $C=C_x=C_y$.

\begin{figure}[h!]
\centering
\begin{subfigure}[b]{0.35\textwidth}
\centering \text{$\vert \mathbf{F}^{\text{Interaction}} \vert$}
\includegraphics[width = 0.9\textwidth]{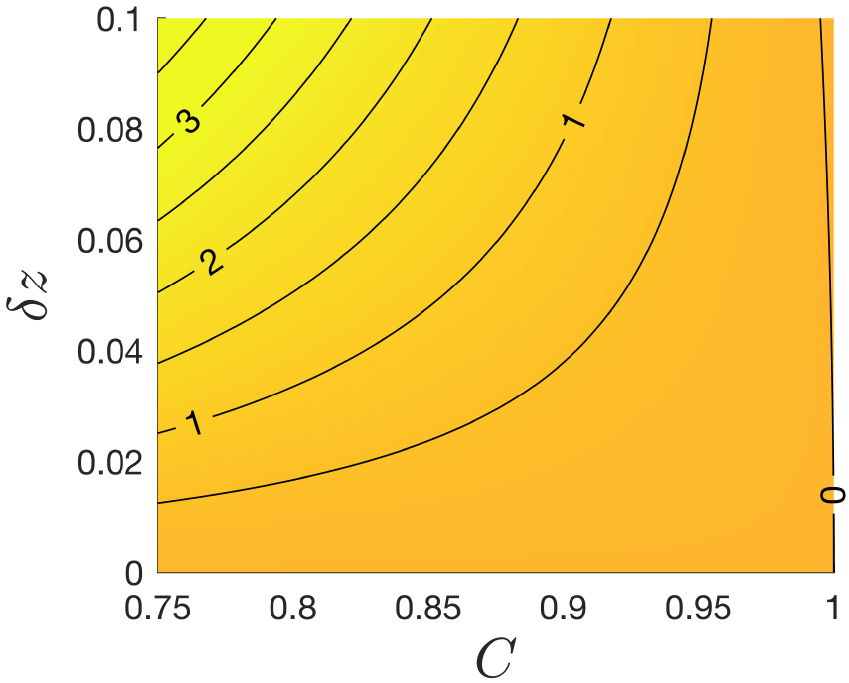}
\caption{\label{fig:model_config_spring}}
\end{subfigure}
\begin{subfigure}[b]{0.35\textwidth}
\centering \text{$-\vert \mathbf{F}^{\text{Bending}} \vert$} 
\includegraphics[width=0.9\textwidth]{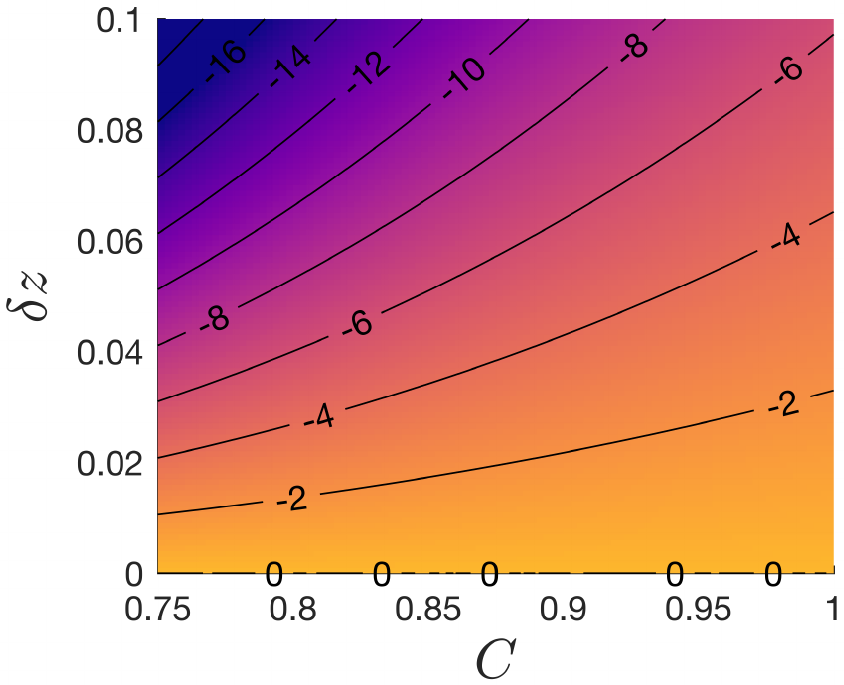}
\caption{\label{fig:bend_inv_bend}}
\end{subfigure}
\begin{subfigure}[b]{0.05\textwidth}
\centering
\includegraphics[width=0.9\textwidth]{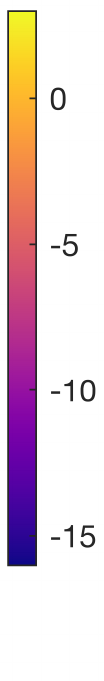}
\end{subfigure}

\captionsetup{subrefformat=parens} 
\caption{ \label{fig:bending_spring} Figure  \subref{fig:model_config_spring} shows the cell-cell interaction force on the centre cell of the test element, with $k_{i} = 20$ and $p_{ij}=1$. Figure \subref{fig:bend_inv_bend} shows the bending force on the same cell with $\rho = 0.2$ and $\alpha = 1.01$. }
\end{figure}

\begin{figure}[h!]
\centering Variation in $\vert \mathbf{F}^{\text{Interaction}} \vert - \vert \mathbf{F}^{\text{Bending}} \vert$ with $\rho$ and $\alpha$ \vspace{0.5cm} 

\normalsize 
\begin{subfigure}[b]{0.02\textwidth}
\centering 
\rotatebox{90}{\text{$\rho = 0.2$}}
\vspace{1.5cm}
\end{subfigure}
\begin{subfigure}[b]{0.30\textwidth}
\centering \text{$\alpha = 1.01$}
\includegraphics[width=\textwidth]{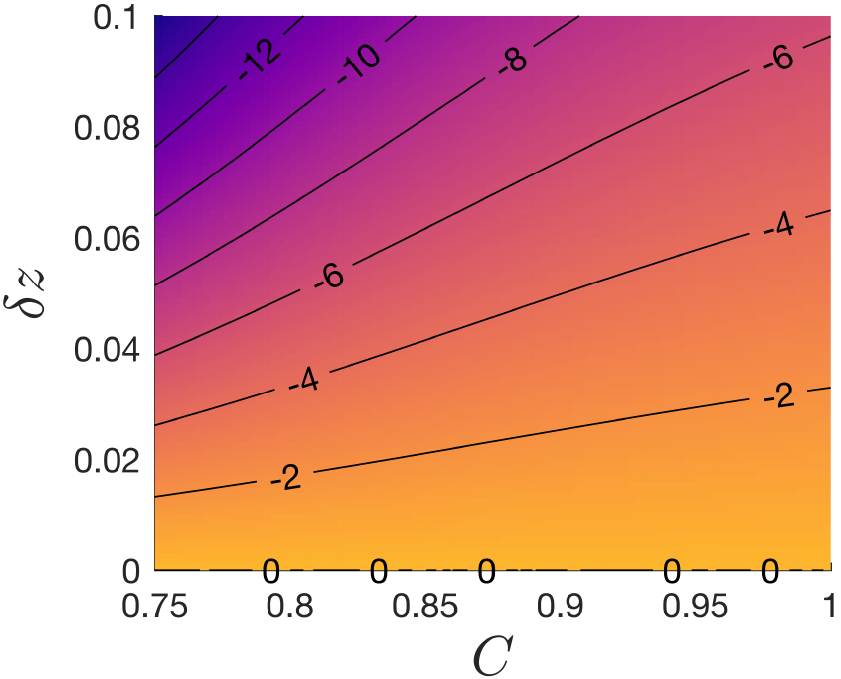}
\caption{\label{fig:net_1p01_0p2}}
\end{subfigure}
\begin{subfigure}[b]{0.30\textwidth}
\centering \text{$\alpha = 1.25$}
\includegraphics[width=\textwidth]{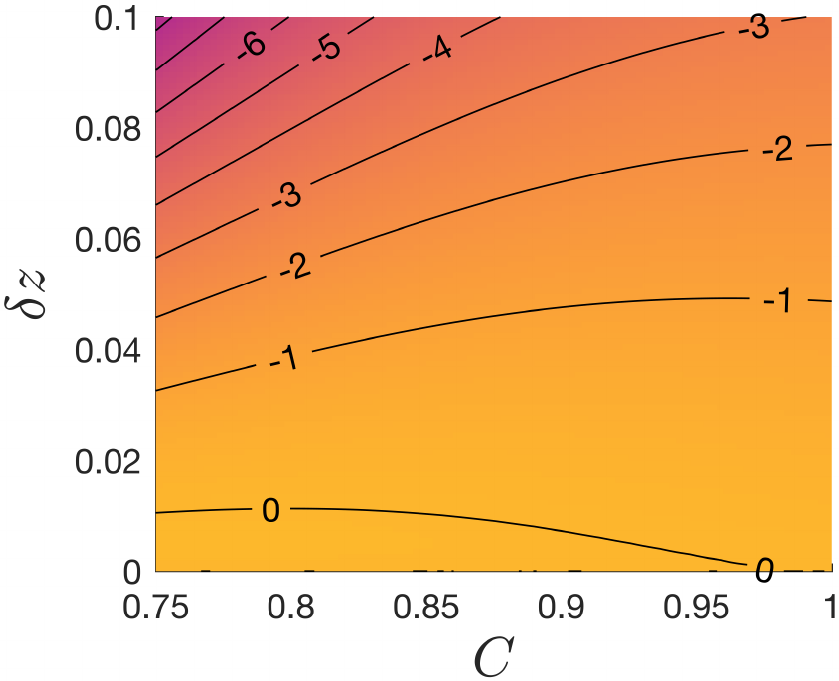}
\caption{\label{fig:net_1p25_0p2}}
\end{subfigure}
\begin{subfigure}[b]{0.30\textwidth}
\centering \text{$\alpha = 1.5$}
\includegraphics[width=\textwidth]{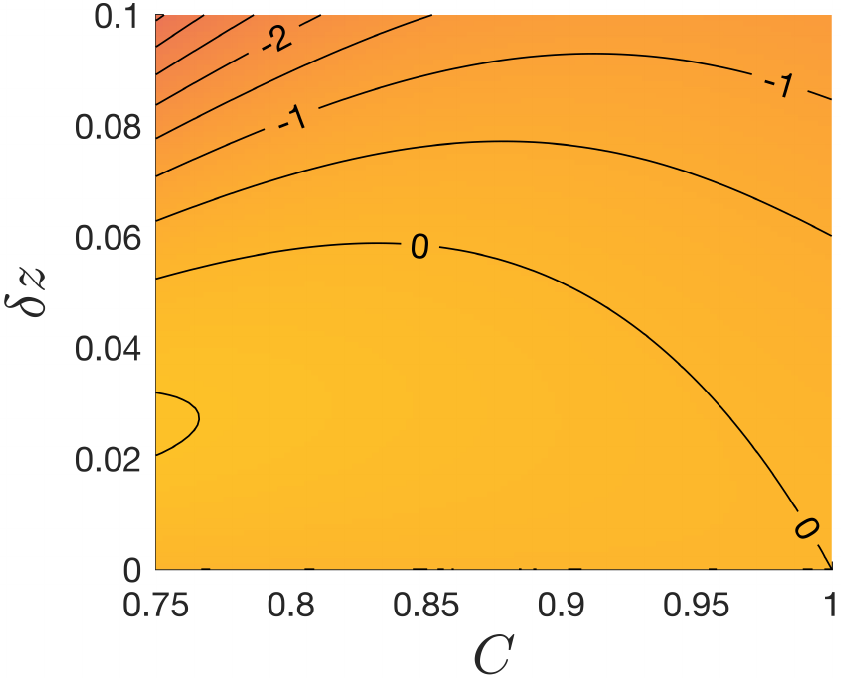}
\caption{\label{fig:net_1p5_0p2}}
\end{subfigure}
\begin{subfigure}[b]{0.0445\textwidth}
\centering
\includegraphics[width=\textwidth]{cbar.pdf}
\end{subfigure}

\begin{subfigure}[b]{0.02\textwidth}
\centering 
\normalsize \rotatebox{90}{\text{$\rho = 0.1$}}
\vspace{1.5cm}
\end{subfigure}
\begin{subfigure}[b]{0.30\textwidth}
\centering 
\includegraphics[width=\textwidth]{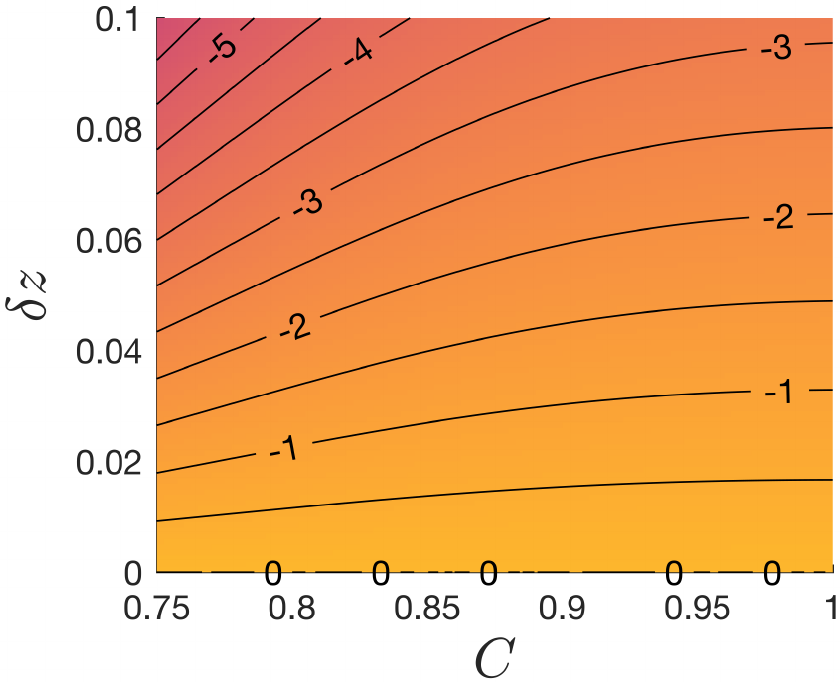}
\caption{\label{fig:net_1p01_0p1}}
\end{subfigure}
\begin{subfigure}[b]{0.30\textwidth}
\centering 
\includegraphics[width=\textwidth]{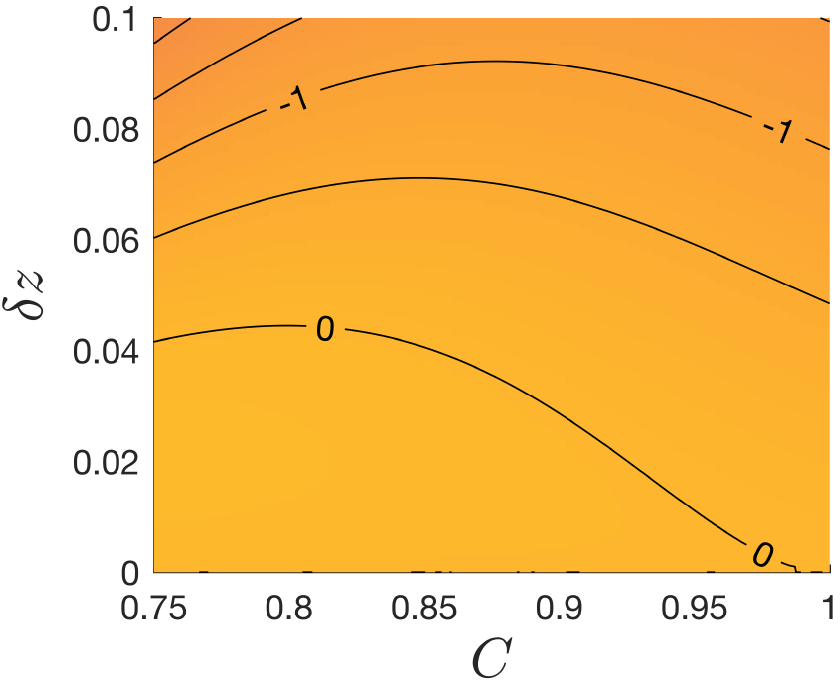}
\caption{\label{fig:net_1p25_0p1}}
\end{subfigure}
\begin{subfigure}[b]{0.30\textwidth}
\centering
\includegraphics[width=\textwidth]{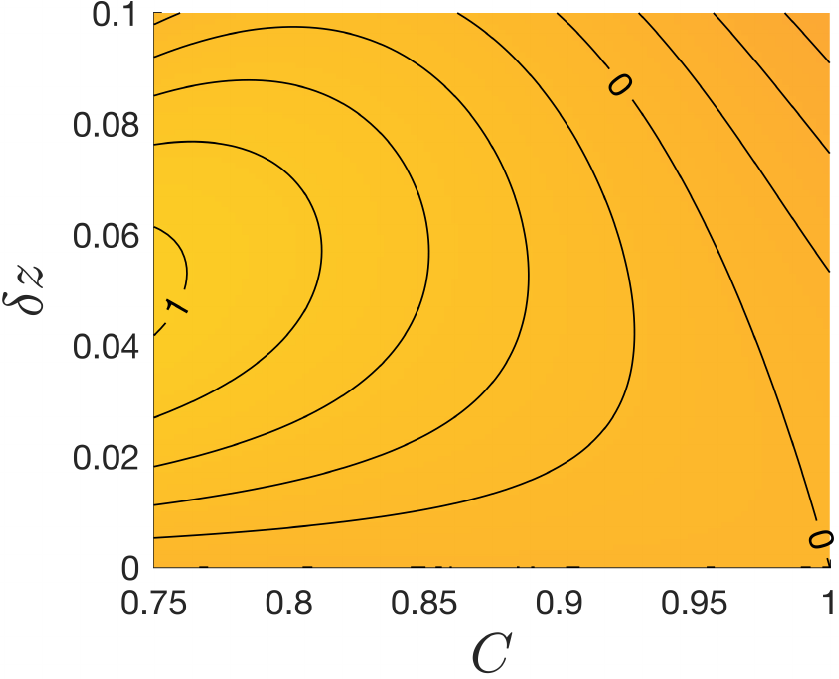}
\caption{\label{fig:net_1p5_0p1}}
\end{subfigure}
\begin{subfigure}[b]{0.0445\textwidth}
\centering
\includegraphics[width=\textwidth]{cbar.pdf}
\end{subfigure}

\captionsetup{subrefformat=parens} 
\caption{ \label{fig:model_config_bend_net} $\vert \mathbf{F}^{\text{Interaction}} \vert - \vert \mathbf{F}^{\text{Bending}} \vert$ for varying values of $\rho$ and $\alpha$.  For $\alpha = 1.01$, \subref{fig:net_1p01_0p1} and \subref{fig:net_1p01_0p2} show the element comes to a near flat configuration. For $\alpha = 1.25$, \subref{fig:net_1p25_0p1} and \subref{fig:net_1p25_0p2} show the element will flatten out, but still come to rest at a height above flat. For $\alpha=1.5$, \subref{fig:net_1p5_0p1} and \subref{fig:net_1p5_0p2} show that as the element is compressed, the cell will always sit above flat, at a height dictated by the zero contour. }
\end{figure}

We specify $k_{i} = 20$ and $p_{ij} = 1$, which gives the interaction force shown in Figure \ref{fig:model_config_spring}, and  the bending force for $\rho = 0.2$ and $\alpha = 1.01$ shown in Figure \ref{fig:bend_inv_bend}. We can find the sum $\vert \mathbf{F}^{\text{Interaction}} \vert - \vert \mathbf{F}^{\text{Bending}} \vert$, as shown in Figure \ref{fig:model_config_bend_net}, and choose $\alpha$ and $\rho$ such that we achieve the desired behaviour of maintaining a flat epithelial monolayer of cells.

Figure \ref{fig:model_config_bend_net} shows  that as we decrease $\alpha$, we control the nullcline which the element will achieve, if we were to move the centre cell only. We need the smallest possible $\alpha$ to maintain a flat layer when the tissue is compressed. However, if we choose $\alpha$ too small, the stability of the numerical solver is compromised. This can be managed by choosing the appropriate $\rho$ to ensure numerical stability is maintained. For this paper, we specify $\rho = 0.2$ and $\alpha = 1.01$.

\section{Video: deformable non-renewing differentiated tissue example}
\label{SI_Movie_1}
The is a simulation sample showing the relaxation of a deformable non-renewing differentiated tissue. Visualised are the epithelial cells (purple - differentiated), the stromal cells, and the triangulation the bending force acts upon.
\url{https://drive.google.com/file/d/1dOSyDGMHbestnN6jFnaYomwx4XfH0497/view?usp=sharing}.

\section{Video: deformable renewing tissue example}
\label{SI_Movie_2}
The is a simulation sample of a deformable renewing tissue. It shows that as the tissue renews, the epithelial monolayer is maintained. Visualised are the epithelial cells (pink - proliferative and red - apoptotic), the stromal cells, and the triangulation the bending force acts upon.
\url{https://drive.google.com/file/d/1aV53ALxFsaygoVmHa4t6WfSULZCl8rcs/view?usp=sharing}.

\section{Video: cell migration in deformable renewing tissue example}
\label{SI_Movie_3}
The is a simulation sample of cell migration in a deformable renewing tissue. It shows that as the cells proliferate in the proliferative region, they then migrate and become differentiated, eventually undergoing apoptosis when they become too compressed. The simulation also demostrates how, even under this process, the tissues structure is maintained.
Visualised are the epithelial cells (pink - proliferative, purple - differentiated and red - apoptotic), the stromal cells, and the triangulation the bending force acts upon.
\url{https://drive.google.com/file/d/1ElUK5eCSEkcpuDflnI3UVBMKaLFLyA5V/view?usp=sharing}.

\end{document}